\documentclass[aps,prd,10pt,twocolumn,superscriptaddress,floatfix,nofootinbib,amsmath,amssymb,preprintnumbers]{revtex4-1}
\usepackage[colorlinks=true,pdfstartview=FitV,breaklinks=true]{hyperref}

\usepackage{bm}
\usepackage{times}
\usepackage{graphicx}
\usepackage{multirow}
\usepackage{soul}
\usepackage{amsmath}
\usepackage{mathrsfs}
\usepackage{amssymb}
\usepackage{yfonts}
\usepackage{color}
\usepackage{xspace}
\usepackage{url}
\usepackage{verbatim}
\usepackage{mathtools}
\usepackage{upgreek}
\usepackage{units}
\usepackage{siunitx}
\usepackage{amstext}
\usepackage{booktabs}
\usepackage{tabulary}
\usepackage{tabularx}
\usepackage{etoolbox}
\usepackage[version=4]{mhchem}
\usepackage[utf8]{inputenc}
\usepackage[dvipsnames,table]{xcolor}
\newcolumntype{Y}{>{\centering\arraybackslash}X}

\hypersetup{urlcolor=BlueViolet,
	    citecolor=Plum,
	    linkcolor=PineGreen}
\usepackage[official]{eurosym}

\definecolor{lightgray}{rgb}{0.9,0.9,0.9}	    
\definecolor{green}{rgb}{0,0.5,0}
\definecolor{red}{rgb}{1,0,0}
\definecolor{blue}{rgb}{0,0,0.5}

\newcommand{\dbd}[2]{\ifmmode \frac{\textrm{d}#1}{\textrm{d}#2}\else $\textrm{d}#1/\textrm{d}#2$\fi}
\newcommand{\pbp}[2]{\ifmmode \frac{\partial#1}{\partial#2}\else $\partial#1/\partial#2$\fi}

\DeclareMathAlphabet{\mathpzc}{OT1}{pzc}{m}{it}
 \newcommand{\eV}{\text{e\kern-0.15ex V}\xspace}

 \newcommand{\TeV}{\text{T\kern-0.1ex \eV}\xspace}

\DeclareMathAlphabet{\mathpzc}{OT1}{pzc}{m}{it}

\newcommand{\be}{\begin{equation}}
\newcommand{\ee}{\end{equation}}
\newcommand{\bea}{\begin{eqnarray}}
\newcommand{\eea}{\end{eqnarray}}

\begin{document}

\title{Particle detection and tracking with DNA}

\author{Ciaran A. J. O'Hare} 
\email{ciaran.ohare@sydney.edu.au}
\affiliation{ARC Centre of Excellence for Dark Matter Particle Physics, The University of Sydney, NSW, Australia}
\affiliation{School of Physics, Physics Road, The University of Sydney, NSW 2006 Camperdown, Sydney, Australia}

\author{Vassili G. Matsos} 
\affiliation{School of Physics, Physics Road, The University of Sydney, NSW 2006 Camperdown, Sydney, Australia}

\author{Joseph Newton} 
\affiliation{School of Physics, Physics Road, The University of Sydney, NSW 2006 Camperdown, Sydney, Australia}

\author{Karl Smith} 
\affiliation{School of Physics, Physics Road, The University of Sydney, NSW 2006 Camperdown, Sydney, Australia}

\author{Joel Hochstetter} 
\affiliation{School of Physics, Physics Road, The University of Sydney, NSW 2006 Camperdown, Sydney, Australia}

\author{Ravi Jaiswar}
\affiliation{School of Physics, Physics Road, The University of Sydney, NSW 2006 Camperdown, Sydney, Australia}

\author{Wunna Kyaw} 
\affiliation{Garvan Institute of Medical Research, NSW 2010, Darlinghurst, Australia}

\author{Aimee McNamara} 
\affiliation{Department of Radiation Oncology, Massachusetts General Hospital, Harvard Medical School, Boston, MA, USA}

\author{Zdenka Kuncic} 
\affiliation{School of Physics, Physics Road, The University of Sydney, NSW 2006 Camperdown, Sydney, Australia}

\author{Sushma Nagaraja Grellscheid} 
\affiliation{Computational Biology Unit, Department of Biological Sciences, University of Bergen, Thormohlensgt 55, Bergen N-5008, Norway}

\author{C\'eline B\oe hm} 
\affiliation{ARC Centre of Excellence for Dark Matter Particle Physics, The University of Sydney, NSW, Australia}
\affiliation{School of Physics, Physics Road, The University of Sydney, NSW 2006 Camperdown, Sydney, Australia}

 \begin{abstract}
 We present the first proof-of-concept simulations of detectors using biomaterials to detect particle interactions. The essential idea behind a ``DNA detector'' involves the attachment of a forest of precisely-sequenced single or double-stranded nucleic acids from a thin holding layer made of a high-density material. Incoming particles break a series of strands along a roughly co-linear chain of interaction sites and the severed segments then fall to a collection area. Since the sequences of base pairs in nucleic acid molecules can be precisely amplified and measured using polymerase chain reaction (PCR), the original spatial position of each broken strand inside the detector can be reconstructed with nm precision. Motivated by the potential use as a low-energy directional particle tracker, we perform the first Monte Carlo simulations of particle interactions inside a DNA detector. We compare the track topology as a function of incoming direction, energy, and particle type for a range of ionising particles. While particle identification and energy reconstruction might be challenging without a significant scale-up, the excellent potential angular and spatial resolution ($\lesssim 25^\circ$ axial resolution for keV-scale particles and nm-scale track segments) are clear advantages of this concept. We conclude that a DNA detector could be a cost-effective, portable, and powerful new particle detection technology. We outline the outstanding experimental challenges, and suggest directions for future laboratory tests.
\end{abstract}

\maketitle

\section{Introduction} 
\begin{figure}[t]
\includegraphics[width=0.49\textwidth]{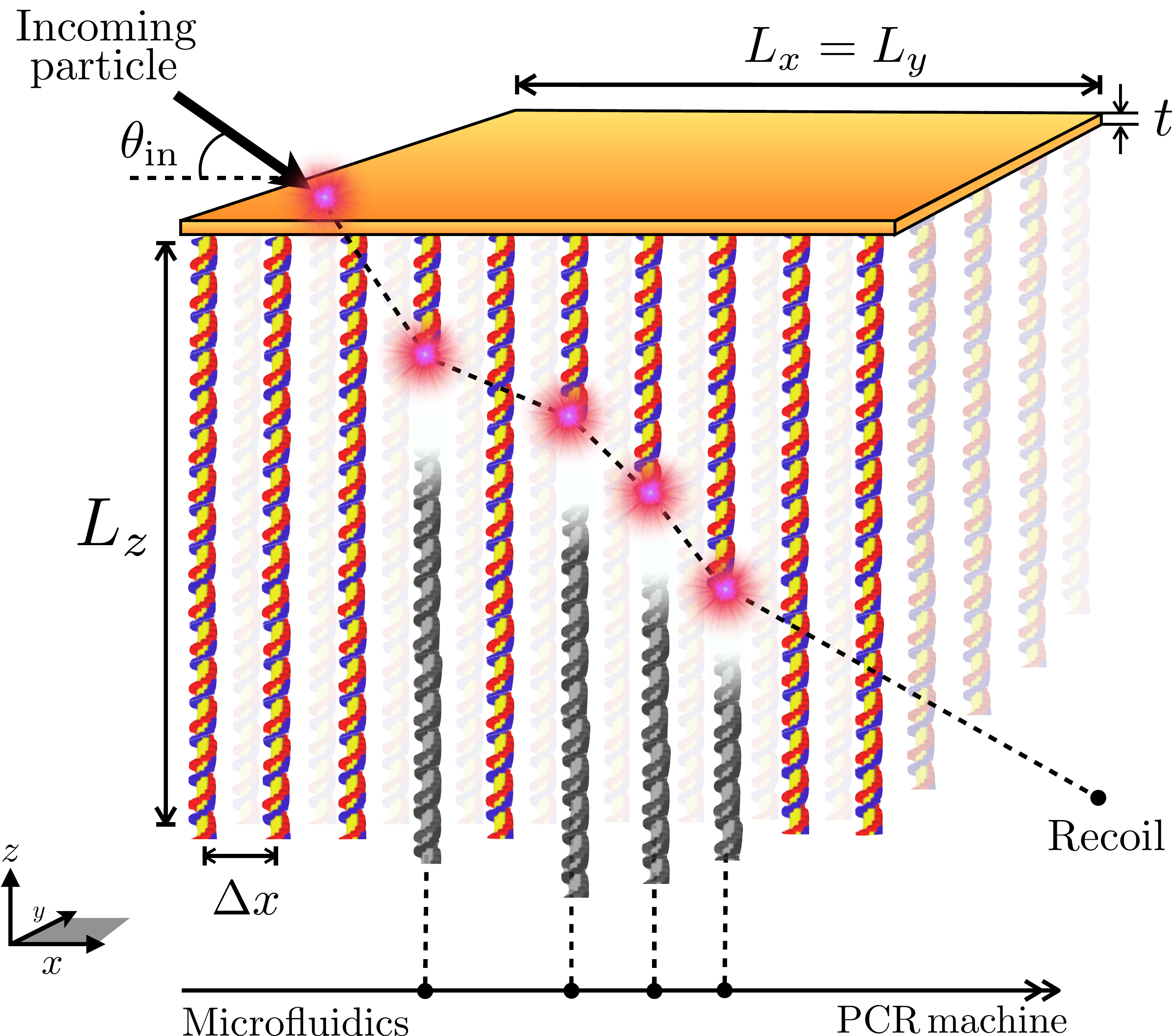}
\caption{\label{fig:diagram} Diagram of the basic concept behind the DNA detector. The detector consists of a thin, dense inorganic holder of area $A = L_x L_y$ and thickness $t$ from which a regular grid of DNA strands are attached with an inter-strand spacing $\Delta x$. The DNA strands, when broken by some incoming particle or secondary recoil, fall down to a collecting area where they are transported in chronological order via microfluidics to an amplification and readout stage consisting of polymerase chain reaction (PCR) devices. A PCR device amplifies the collected strands to where the precise sequence of bases can be measured and therefore each vertex in the track can be reconstructed.}
\end{figure}

In the search for signatures of new physics, fresh experimental ideas are always desired. Ultrahigh energy cosmic rays~\cite{ThePierreAuger:2015rma}, weakly-interacting particles orbiting the galaxy~\cite{Battaglieri:2017aum,Schumann:2019eaa,Undagoitia:2015gya}, light species emitted by the Sun~\cite{Raffelt:1999tx,vanBibber:1988ge,Anastassopoulos:2017ftl,Redondo:2013lna,Aprile:2020tmw,Irastorza:2018dyq}, or even the search for relics of the Big Bang~\cite{Baracchini:2018wwj}, are just a few of the frontiers in particle physics which would greatly benefit from new technologies. Recently there has been substantial activity in the community towards the repurposing of interdisciplinary science. Ideas leveraging cutting edge research in condensed matter physics~\cite{Hochberg:2015fth,Hochberg:2015pha,Knapen:2016cue,Budnik:2017sbu,Hertel:2018aal,Marsh:2018dlj,Schutte-Engel:2021bqm,Griffin:2020lgd}, nuclear magnetic resonance~\cite{Budker:2013hfa,Crescini:2016lwj}, nanomaterials~\cite{Cavoto:2016lqo,Cavoto:2017otc,Hochberg:2019cyy} and even microscopy of ancient rocks~\cite{Goto:1963zz,Fleischer:1970zy,Kovalik:1986zz,Collar:1994mj,SnowdenIfft:1997hd,SnowdenIfft:1995ke,Drukier:2018pdy,Baum:2018tfw,Ebadi:2021cte,Acevedo:2021tbl} have all been proposed in the context of new particle detection techniques.

This study follows a similar spirit, but instead seeks to use technologies from the biosciences. From proton therapy to medical imaging, particle physics techniques have repeatedly proven their use in biological applications since the 1950s~\cite{Dosanjh}. Here we propose to return the favour, by asking how we might leverage advances in biotechnology to answer fundamental questions in particle physics. 

The concept of a DNA-based particle detector was originally proposed in articles by Drukier et al.~\cite{Drukier:2012hj,Drukier:2014rea} to help elucidate the answer to one of those questions: The nature of dark matter---a mysterious substance that is supposed to fill the galaxy in which we live~\cite{Bertone:2016nfn}. The basic detector concept is shown in Fig.~\ref{fig:diagram}. We follow the essential features of the original design of Ref.~\cite{Drukier:2012hj}, but incorporate some additional features that we will discuss shortly. The general idea consists of strands of DNA attached to a thin gold sheet. In the original proposal, dark matter particles from our galaxy would enter the detector, scatter primarily off of the gold nuclei in the foil, and send them on recoils through the DNA forest, breaking a sequence of strands in a characteristic pattern. The broken segments would be encouraged to fall down to a collecting area with the use of magnetic beads attached to their ends (the beads would also act to keep the DNA strands approximately straight while in the detector). After that, they would be transported to devices that perform polymerase chain reaction (PCR) to amplify the strand segments. Then, by studying the precise sequences of bases in the collected strand segments, one could infer the original spatial positions in the detector at which each strand was severed, thereby reconstructing complete particle tracks.
 
Since the original DNA detector proposal, state-of-the-art dark matter detectors~\cite{Undagoitia:2015gya,Schumann:2019eaa,Battaglieri:2017aum} have exceeded the ton-scale in size---providing a level of sensitivity that is difficult for entirely new technologies to compete with. Therefore the primary benefit of such a DNA detector in the context of dark matter right now would be the ability to perform \emph{directional} measurements of low-energy recoils. Indeed, dark matter signals are expected to be strongly directional~\cite{Mayet:2016zxu}, a phenomenon generated by the orbit of the solar system through the dark matter halo that envelopes our galaxy~\cite{Spergel:1987kx}. A search for recoil tracks aligning with the direction of our galactic rotation would permit a convincing test of the veracity of any potential dark matter signal~\cite{Mayet:2016zxu,Morgan:2004ys,Green:2007at,Billard:2009mf,Green:2010zm,Billard:2011zj,Vahsen:2021gnb}, but would also allow it to be cleanly distinguished from sources of background noise such as cosmic rays, radioactive decays, and neutrinos~\cite{Billard:2014ewa,Grothaus:2014hja,O'Hare:2015mda,OHare:2020lva}.

The reconstruction of particle tracks is challenging at the low energies relevant for a dark matter search. Hence, directional information is usually not sought after in the most competitive experiments, instead requiring more specialised detectors~\cite{Sciolla_2009,Ahlen:2009ev,Battat:2016pap}. After many years of steady progress by smaller collaborations~\cite{Santos:2011kf,Baracchini:2020btb,Battat:2016xxe,Yakabe:2020rua,Vahsen:2011qx}, only very recently have the steps been laid out for a directional dark matter detector at a competitive scale~\cite{Vahsen:2020pzb}. Even though directionality is achievable in several classes of detector, such as time projection chambers, these are not without limitations (see the discussion in Ref.~\cite{Vahsen:2021gnb}). So given the great benefits brought by directional sensitivity, both for dark matter physics~\cite{Lee:2012pf,Billard:2010jh,O'Hare:2014oxa,Catena:2015vpa,Kavanagh:2015jma,Kavanagh:2016xfi,OHare:2018trr,OHare:2019qxc}, and in many other contexts~\cite{Vahsen:2021gnb,Leyton:2017tza,nuBDX-DRIFT,Baracchini:2020owr}, new directionally sensitive technologies will always be welcome. A DNA-based detector could therefore represent a major breakthrough in the field if it were demonstrated to be practical.

Yet even if the concept proved to be too impractical at a scale large enough for a dark matter search, there is still value in having new directional particle detector technologies. Directional detectors at all scales facilitate a wide array of fundamental and applied physics measurements~\cite{Vahsen:2021gnb}. Therefore, a DNA-based detector, which could have both impressive directionality and an extremely fine spatial granularity, could eventually serve an even wider array of applications. In this paper, therefore, we aim to simulate how a DNA detector would respond to a range of particle types. These particles could be the nuclear recoils generated by neutrons or dark matter scattering with the foil holder, but also charged cosmic rays impinging on any part of the detector from any direction, as well as on the gold foil.

To begin, in Sec.~\ref{sec:motivation} we explain the rationale behind pursuing this kind of detector, and explain its potential novelty. Section~\ref{sec:setup} presents details of the setup of the DNA experiment and various practical considerations which will inform how we must configure our simulation. In Sec.~\ref{sec:simulation} we describe this simulation, its various inputs, and ways in which we can measure events and tracks. In Sec.~\ref{sec:results} we present our results, and in Sec.~\ref{sec:outlook} we outline several questions that will require further experimental investigation. We conclude in Sec.~\ref{sec:conclusions}.

\section{Rationale}\label{sec:motivation}
In most of the widely-used low-energy particle detectors, even a hypothetical reconstruction of three-dimensional recoil track vectors is difficult to imagine. In dedicated directional detectors, like gas-based detectors, the solution is to drift the entire mm-scale ionisation cloud generated by recoiling nuclei or electrons to a segmented readout plane. This transportation inevitably causes diffusion of the initial track topology. Hence for short tracks, long drift lengths, or high-density fill gases, initial recoil directions can often be washed out, especially at low energies.

Alternative methods of directional detection that rely on detecting tracks lodged in solid-state targets~\cite{Agafonova:2017ajg,Marshall:2020azl,Barreto:2011zu,Crisler:2018gci,SnowdenIfft:1997hd,SnowdenIfft:1995ke,Drukier:2018pdy,Baum:2018tfw} can circumvent diffusion, however these typically require challenging nanoscale imaging to obtain good directional sensitivity. Furthermore, many of these approaches cannot be performed in real time, so no event-by-event timing information is available. The Earth's rotation washes out the directionality of any astrophysical source if events are not tagged in time~\cite{OHare:2017rag}.\footnote{We note, however, that the dark matter detector proposals of Ref.~\cite{Agafonova:2017ajg} and~\cite{Marshall:2020azl} do both suggest methods of either reclaiming the event time information or mitigating against the lack of it.}.

Although there are many other approaches for detecting low energy recoil directions that do not rely on direct imaging of tracks, see e.g. Refs.~\cite{Hochberg:2021ymx,Nygren:2013nda,Cappella:2013rua}, the DNA detector would still have several interesting characteristics. Most notably it would provide nm-scale spatial reconstruction of recoil tracks whilst simultaneously circumventing diffusion. To fully augment this technology's capabilities to study even astrophysical particle sources, we only need a method of reclaiming timing information. This is why, as shown in Fig.~\ref{fig:diagram}, we suggest that the detector should be fitted with a system of microfluidics that would act as a conveyor-belt to transport the strands chronologically to the PCR amplification stage. Processing the severed strands in batches every few hours would be sufficient to counteract the rotation of the Earth~\cite{Vahsen:2021gnb}. See e.g.~Ref.~\cite{Kam2013} for a previous study using real-time PCR to quantify radiation-induced DNA damage.

To achieve the claimed nm-precise track measurements, nucleic acid base pair information must be arranged and reconstructed on the same nm scales. While this may seem challenging, this requirement is already met by current bio-engineering and DNA sequencing and is easy to achieve. The main question that needs to be addressed is: How do ionising particles interact with DNA when arranged in this way? DNA damage induced by different types of ionising radiation has been the subject of extensive research over many decades~\cite{Lea_radiation,Chadwick_radiation,Goodhead1989,Nikjoo2016,Mavragani_2019}. Our study, on the other hand, asks whether DNA-strand breakage occurs readily and predictably enough to be a basis upon which to design a particle detector. Answering this question will also allow us to determine the particle types and energy scales for which such a concept could be applied. 

Currently, the basic DNA detector concept shown in Fig.~\ref{fig:diagram} has no theoretical or experimental validation, but it has many attractive properties that warrant performing a proof-of-concept investigation with simulations. As well providing time-resolved 3-d particle directions, the detector, in principle, could be manufactured cheaply, with commercially available components, and operated with widely-used biotechnology techniques.

Ultimately, our aim is to make the first quantitative analysis via simulation of the heuristic arguments proposed in Ref.~\cite{Drukier:2012hj}. To achieve this, we perform Monte Carlo simulations based on the widely-used particle interaction software Geant4~\cite{Geant4_1,Geant4_2} using basic DNA-strand models constructed with the Geant4 front-end software TOPAS~\cite{TOPAS} and the TOPAS-nBio radiobiology toolkit~\cite{TOPASnBio}.

\section{Detector setup}\label{sec:setup}
A large number of factors would need to be taken into consideration when constructing a detector with a level of complexity like this. Therefore, as a starting point, we follow the baseline configuration proposed in Ref.~\cite{Drukier:2012hj} where the detector is comprised of a holder made of a thin film of high-density material (in their case gold), from which long strands of single or double-stranded nucleic acids are hung. As shown in Fig.~\ref{fig:diagram}, it is assumed that a particle (either a primary particle incident on the DNA, or a secondary product of an interaction taking place elsewhere in the detector) will break a sequence of strands in a roughly straight line that correlates with its direction of travel. 

We begin by orienting a Cartesian coordinate system where the $x-y$ plane is parallel to the holder, and the $z$ direction is parallel to the strands. The resolution in the plane is given by $\Delta x$, which we will refer to as the interstrand spacing. The total volume of the detector is given by $V = A \cdot L_z$ where $A$ is the area of the holder, and $L_z$ is the length of each strand. Before placing the strands, the box is a vacuum. Ideally, we would simulate detectors with side lengths larger than $\sim \upmu$m. However, due to rapid increases in computational cost in building a detector event counter with such a high spatial granularity we are limited in the overall size we can simulate to around 10 $\upmu$m$^3$. Furthermore, due to the low density of the detector, it will generally be the case that incident particles are unlikely to be stopped completely inside $V$. This is also likely to be true of a real experiment which will probably need to consist, for practical reasons, of several smaller detector modules. While there should be many additional observables that can be extracted from particle tracks extending across multiple smaller units, we defer this kind of analysis to a future study building on the present one. Here we seek to understand the signals present at the nanoscale, since that is one of the claimed advantages of this concept.

We now briefly discuss the important details of the principal components of the detector: the strands, the holder, and the readout.

\subsection{The strands}
\begin{figure}[t]
    \includegraphics[width=0.49\textwidth]{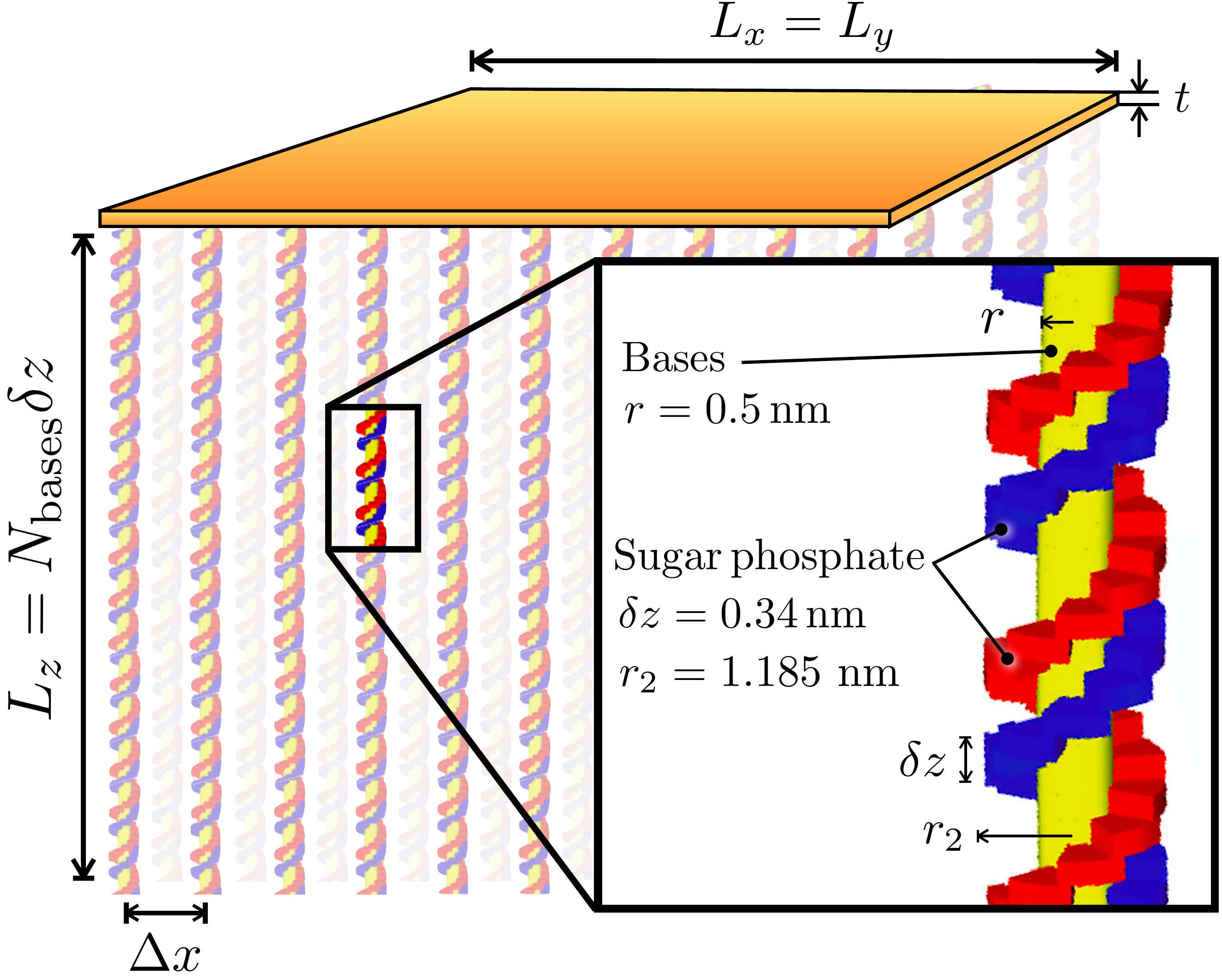}
    \caption{\label{fig:DNA_model} Closeup of DNA strand model setup constructed in TOPAS and which is used to count incoming particle interactions with Geant4. The model shown here is a simplification of a double stranded DNA molecule. The central yellow cylinder with radius 0.5~nm represents the sequence of base pairs, which for simplicity we assume are all composed of adenine. The winding sugar phosphate backbones (red/blue) each are made of cylindrical segments with heights in the $z$-direction of 0.34~nm. Since we assume a 2000 base-long DNA strand, the total height of the detector $L_z = 680$~nm. Though this is not in any way a biologically accurate model of a DNA strand, it does still reflect the basic geometry and broad composition, which is what is important for our simulation.}
\end{figure}

Each strand of DNA is built individually using the model shown in the closeup of Fig.~\ref{fig:DNA_model}. A more realistic model of a DNA strand would be prohibitively complex for these simulations, but it is also not obvious that it would change any of our results substantially. Instead, we use this simplified model to represent the basic composition and geometry. Each strand consists of a central cylinder representing the bases, with either one or two sugar phosphate backbones made of quarter cylinders wrapping around the bases. We find no quantitative difference in results by using single or double-stranded models up to an increase in interaction rate for the latter, which we utilise in all later results. We fix a total strand length of 680 nm, which corresponds to 2000 of these bases. This represents a long, but computationally feasible length that would also be around the longest strand that could be sequenced cheaply in the lab with naturally-occurring DNA, or PCR by-products. The bases are crudely modelled using only a single nucleobase, adenine, with molecular ratio 4:5:5 (H:C:N) and a mean excitation energy of 72~eV, whereas the backbone is modelled as a custom material with a 2.26:0.44:5.59:1.35 (C:H:N:P) molecular ratio and the same mean excitation energy.

An essential assumption we must make is how much energy deposition is needed to break a DNA molecule. 
Experimental and theoretical studies, e.g.~Refs.~\cite{DNAdamage-expt,DNAdamage-theory}, show that a DNA strand can be broken by ionising radiation depositing energies above 10 eV. In principle, this energy scale sets the absolute minimum threshold for registering a strand breakage, though other energy scales required for detection will almost certainly be larger than this. So we implement 10 eV as a minimum threshold for registering a strand break site, although it is generally the case that the majority of primary interactions deposit sufficient energy.

To get the strands to fall to the readout plane, it is envisaged that the free end of each strand that is not attached to the holder would be fitted with a magnetic bead. The magnetic pull would then act to keep the strands in place, prevent them from curling up~\cite{BustamanteDNATension}, and would attract the severed segments towards the bottom of the detector to be collected. Several complications may arise with this strategy, for instance the strands are not guaranteed to be absolutely straight and hence an error in the precise $z$ position of each base will inevitably be incurred. We will discuss this issue and others in Sec.~\ref{sec:noise}. But for now, we simply need to specify a value for $\Delta x$. Ensuring that the strands fall and arrive at the readout will place a restriction on $\Delta x$ since we must ensure that, subject to the magnetic pull, the strands fall to the readout plane without connecting to their neighbouring strands in some way. We also need to ensure that the strands can be attached to the holder at regular intervals of $\Delta x$. Since this optimisation will require experimental investigation we do not vary $\Delta x$ from the baseline value of 10 nm suggested by Ref.~\cite{Drukier:2012hj}. Importantly, for the small-scale simulations we employ here, varying $\Delta x$ does not meaningfully change our results. Since the energy depositions, $\textrm{d}E/\textrm{d}x$, are still relatively low, a larger value of $\Delta x$ merely amounts to a rescaling of the total track lengths by the same factor. We also do not explore non-rectangular grids for the strands, though this is also something that could be investigated and would enhance the density of the detector.

\subsection{Holder}
In Ref.~\cite{Drukier:2012hj} the holder layer was assumed to be made from gold, and for the sake of having a benchmark, we use this material in our simulation as well. However, we note that gold is not the only feasible choice for the holder material, and the majority of our results either do not depend on this choice, or could be qualitatively mapped on to any other relatively dense material. 

The main requirements of the holder material are: 1) that it provides a high-density source of nuclear recoil events, 2) that it can be manufactured to nm-scale thicknesses, and 3) that it can function as a substrate for the attachment of DNA. A more informed selection of the holder material would therefore require more optimisation, and this would depend upon a specific purpose to be chosen for the detector. We outline how such an argument could be made below.

The first requirement is to enhance the rate of nuclear recoils, which would be particularly attractive in the context of searches for dark matter in the form of weakly interacting massive particles. These searches benefit from heavier nuclei, since spin-independent dark matter cross sections would scale with the mass number squared when the DM interacts coherently across the nucleus (see e.g.~Ref.~\cite{Cerdeno:2010jj}). So while gold is certainly attractive from this point of view, such a requirement may not be so necessary for all applications, like for studying cosmic rays or radioactive backgrounds. Rather, the consequently higher density could be detrimental. A rough Lindhard-Ziegler stopping power calculation~\cite{SRIM} suggests that a 10 nm thickness of gold foil would generally stop the majority of nuclear recoils below around 20 keV. Even for recoils scattering slightly above this, the typical final energies will be low and harder to detect, and the scattering angles would be much wider than the angle of the initial recoil generated inside the foil. Therefore a good balance for other applications may be a lower density material like copper. Copper is a commonly used material in particle detectors requiring low background materials. Screening of copper for the NEXT-100 neutrinoless double beta decay experiment, for example, revealed ultralow-background levels of $\lesssim$0.08 mBq/kg (U, K and Th)~\cite{Alvarez:2012as}. High-purity copper nanofoils with thicknesses down to around 10~nm are commercially available. However, as the holder makes up the majority of the target mass of the detector, a highly \emph{radiopure} material is likely to be non-negotiable. For instance, ideally this copper would be stored in underground conditions as much as possible to avoid cosmogenic activation~\cite{Cebrian:2017oft} of $^{60}$Co. Ultimately, while this strategy may be reasonable in the context of copper, it may be more difficult to achieve with gold foils.

The final major requirement of the holder material mentioned above is to act as a substrate to attach the strands to. This is relatively common practice in DNA nanotechnology. Another motivation behind gold in this context is that gold-DNA attachment is readily achievable via the use of a sulfur-based thiol group~\cite{Liu_AuNP}. The experimental challenge that this presents is how to ensure the strands are arranged in a regular pattern with precisely known positions. Several options are available, but will need experimental investigation. One could be to use geometric nanowells arranged in a zig-zag like pattern to create preferential positions for the thiol attachment, as in Ref.~\cite{Visnapuu}. Alternatively, the substrate could be pre-prepared with DNA origami nanostructures to which the strands would then attach~\cite{Sajfutdinow,Busuttil}. 

For the detector to achieve its claimed nm-precise position reconstruction, the positions at which strand are placed need to be known to the same precision. We note in passing however, that is not the same as requiring that the strands are placed at specific locations to nm-precision: it is our \emph{knowledge} of where the strands are that needs to be nm-precise. Therefore, a setup in which a substrate was prepared with DNA origami structures arranged roughly in a grid, then precisely imaged using some form of microscopy (e.g. the nm-precise technique called DNA-PAINT~\cite{DNA-PAINT}), would work just as well.

For our simulation, we make no assumptions about exactly how the strands are attached to the holder. We simply simulate the holder as a square layer of gold with a constant density of 19.32 g~cm$^{-3}$, area $A = 1\upmu$m$^2$, and thickness $t = 10$~nm.

\subsection{Readout}

The precise readout and reconstruction process is not the subject of this study, however we make a few comments to add some clarity about how this stage could work in theory. In broad terms, the envisioned readout will consist of a layer using microfluidics to transport the fallen strands in a chronological order out of the detector volume. Batches of strands would then be amplified and read-out using a host of PCR machines.  

PCR machines are widely used across many fields, from biomedical research to archaeology, whenever DNA samples are needed to be amplified to detectable levels. A conventional PCR machine performs a sequence of chemical processes which, when repeated many times, results in the exponential replication of an initial input DNA strand template. There are three basic stages to PCR: first, the DNA sample is heated to 95--98$^\circ$C causing it to denature into its single strands; then, the sample is cooled to allow specially chosen complementary `primer' strands to anneal to the ends of the single strands; then in the final step, an enzyme called DNA polymerase is used to elongate the annealed strands resulting in two copies of the initial template DNA. The process can then be repeated on the copied templates to achieve exponential multiplication of the DNA. To optimise the process, the sample must be heated and cooled to facilitate the three temperature-dependent reactions. This is achieved with an instrument known as a thermocycler which has recently been adapted for use in microfluidic devices~\cite{Ahrberg2016}.

The PCR machine would effectively act as an amplification stage (in analogy to a gas electron multiplier or other similar technologies used in conventional directional detectors) and will allow the sequence of bases that make up each strand to be reconstructed. The main limitation here is on the length of the DNA strands $L_z$ that can be amplified both efficiently and accurately. This will limit the maximum dimension of the detector unit in the $z$ direction. We fix $L_z$ to 680 nm in this study which corresponds to around 2000 bases and a total volume of $0.69~\upmu$m$^{3}$. We anticipate that a full-size detector would consist of a coordinated array of these smaller units. The PCR output can then be sequenced using high throughput approaches, such as Illumina or Nanopore sequencing, as long as the limitations of these in terms of accurate reconstruction of the precise strand sequences are do not too severely limit the resolution in $z$. We discuss this latter issue in more detail in Sec.~\ref{sec:noise}.

\subsection{Event reconstruction}\label{sec:eventrecon}
After the readout stage and the sequences of each broken strand have been reconstructed, this information must then be converted back into the $(x,y,z)$ positions of each break site. To reconstruct a track from breakage sites both the initial locations of each strand on the holder $(x,y)$, and the particular base at which each strand was cut must be measured precisely. It is relatively safe to assume that once the sequence of bases of each cut strand is known (via amplification at the PCR stage) the three coordinates can be determined precisely. This must be enabled by sequencing each individual strand in the detector with a unique, and precise, pattern of bases. Algorithmically sequencing the strands to cover even large volumes is not anticipated to be problematic. For example, a particular terminal sequence of bases appended to the very end of the strand (before the magnetic bead) would mark the $x-y$ position, and then the rest of the $z$-extent of the strand could be identical throughout the whole detector. A combination of many marker sequences could then be used to arrange the strands inside each smaller unit to construct a larger detector. 

In the current detector design, we can only assume that each strand is able to detect a single interaction. Once a strand is broken by an interaction, the part of the strand that remains attached to the holder will no longer be stretched out by the magnetic force which kept the bottom end in place prior to the interaction. Therefore it is highly likely that this strand segment will begin to curl up. While it is possible for an interaction to sever part of the curled strand, it may be undesirable to use this information in analysis because the reconstructed bases would no longer correspond to the $z$-position of the interaction. Ultimately this means that the detector has a finite lifetime, with an effective target mass that diminishes with the number of interactions it measures. Determining this lifetime would require additional input that is outside of the focus here: including the background conditions of the detector, as well as the typical exposure that would be needed for some particular science goal.

An added benefit of nm-scale position reconstruction within the detector, is that this style of experiment is naturally bestowed with an obvious strategy for fiducialization. This refers to the ability to define a volume within the interior of the detector bulk where the rate of background events is expected to be significantly lower. This is typically because rare or non-background events are equally likely to scatter anywhere in the detector volume, whereas background events, e.g. from radioimpurities, are more likely to occur close to the edges of the detector. In many dark matter experiments, for instance, fiducialization of the detector volume is an important design consideration~\cite{Lewis:2014poa}. Especially in gas-based directional detectors, which drift tracks of ionisation from the point of deposition to the readout, absolute position localisation along the drift direction often places requirements on the type of gas and readout used~\cite{Snowden-Ifft:2014taa,Phan:2016veo}. As mentioned above, we do not perform any background studies here, but simply remark that the precise position reconstruction brings additional advantages that are not the focus of this work.

\begin{figure}[t]
    \includegraphics[width=0.49\textwidth]{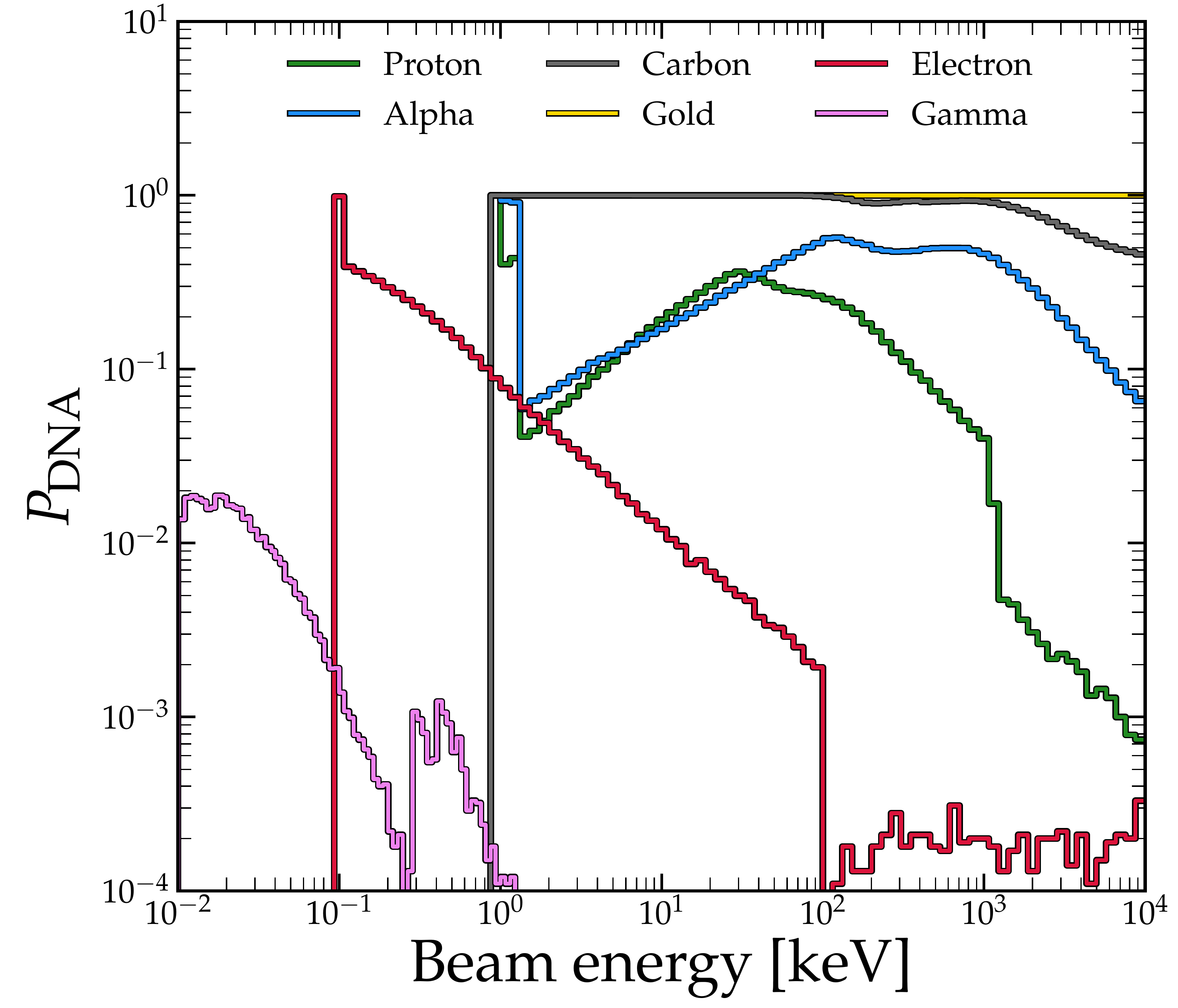}  
    \caption{\label{fig:PDNA} Probability of particle-DNA interaction depositing an energy larger than 10 eV in a single component volume of the DNA, when the beam was fired directly at a strand. This is a proxy for the particle-DNA cross-section for severing a strand. The spurious threshold effects around 0.1 keV and 100 keV for electrons and 1 keV for ions are due to limitations of the available electromagnetic physics models. Hence we will generate or study incident particle energies outside of these ranges where we are not able to simulate the complete physics.}
\end{figure}

\begin{figure*}[t]
    \includegraphics[width=0.49\textwidth]{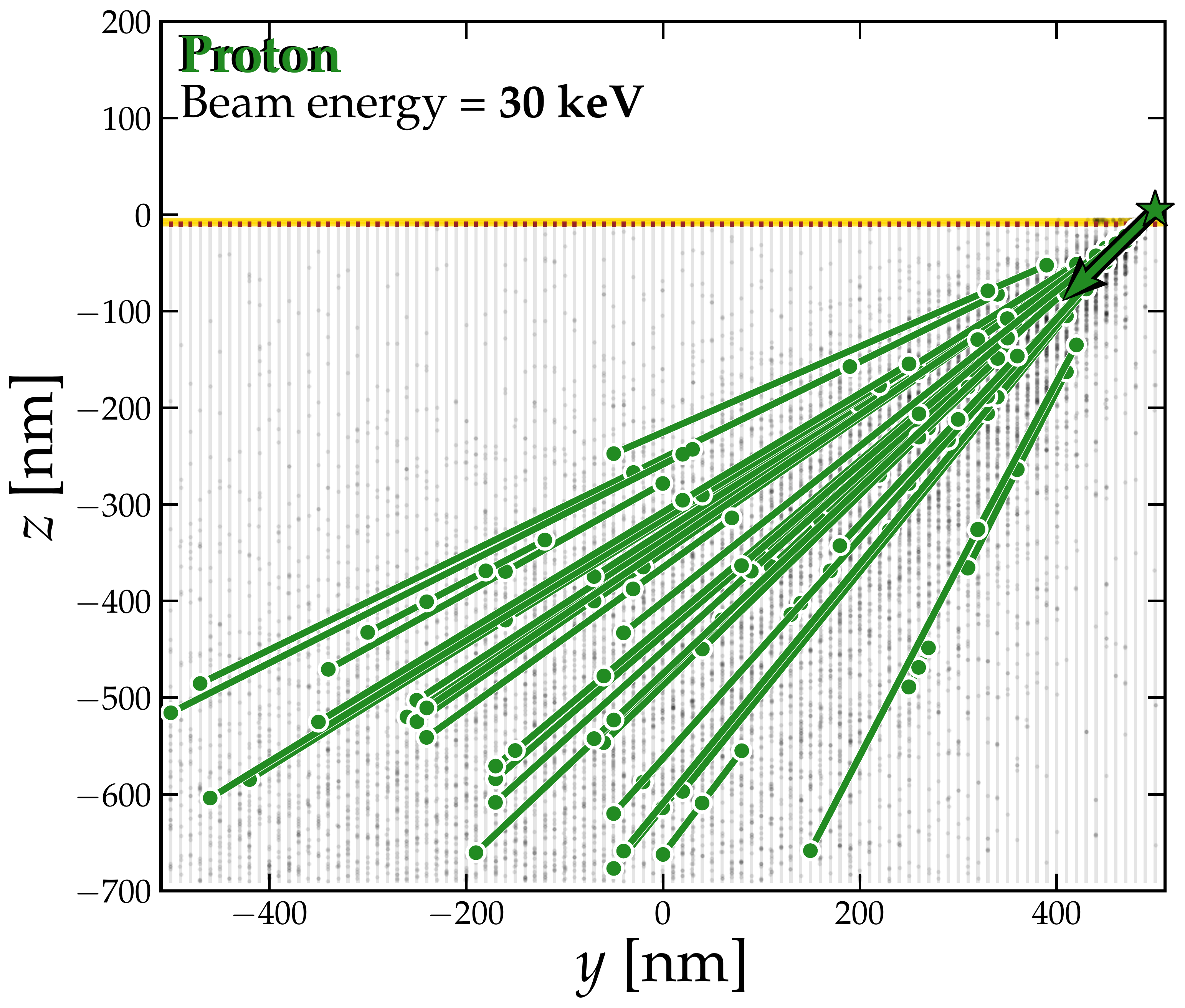}    \includegraphics[width=0.49\textwidth]{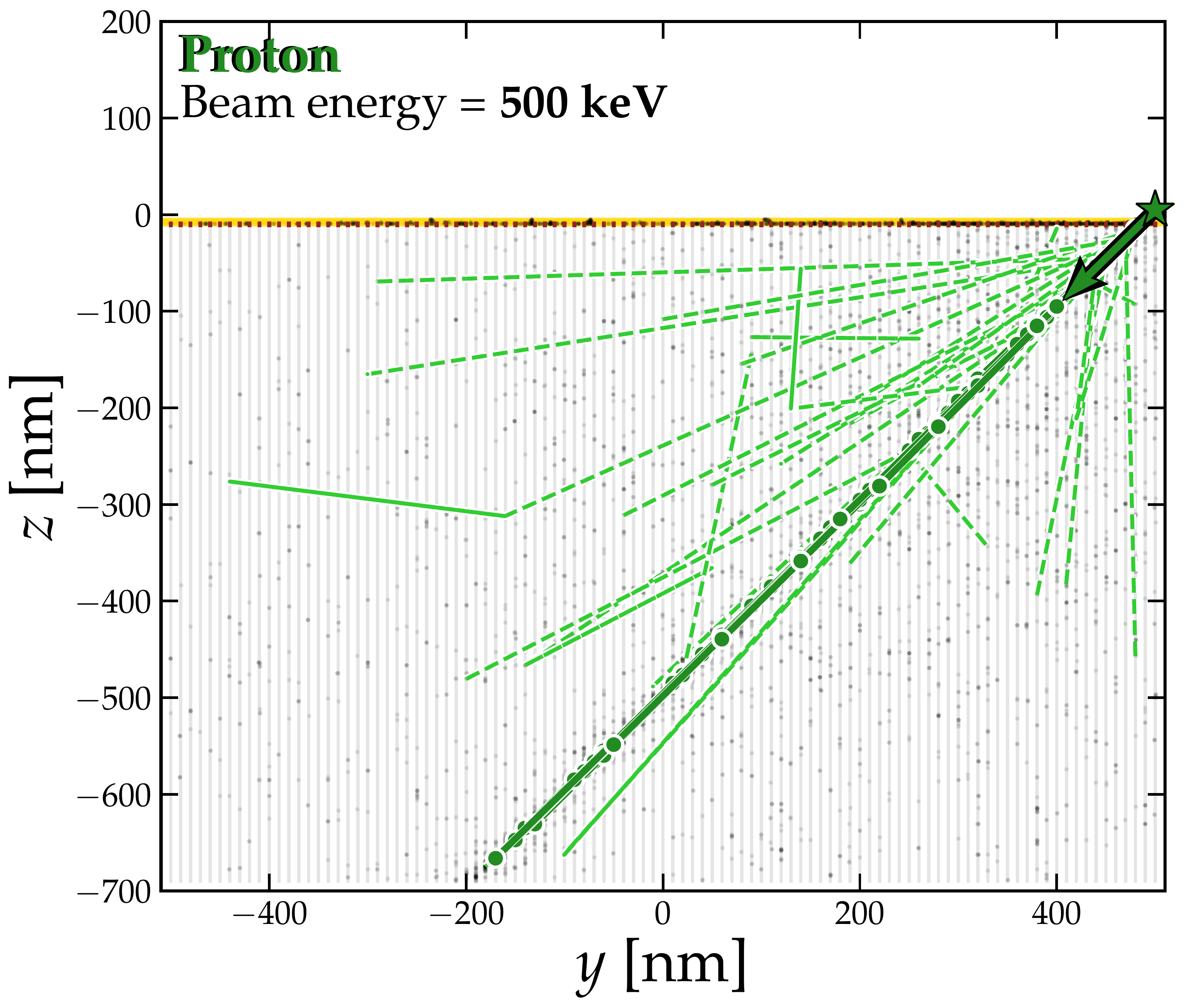} 
    \includegraphics[width=0.49\textwidth]{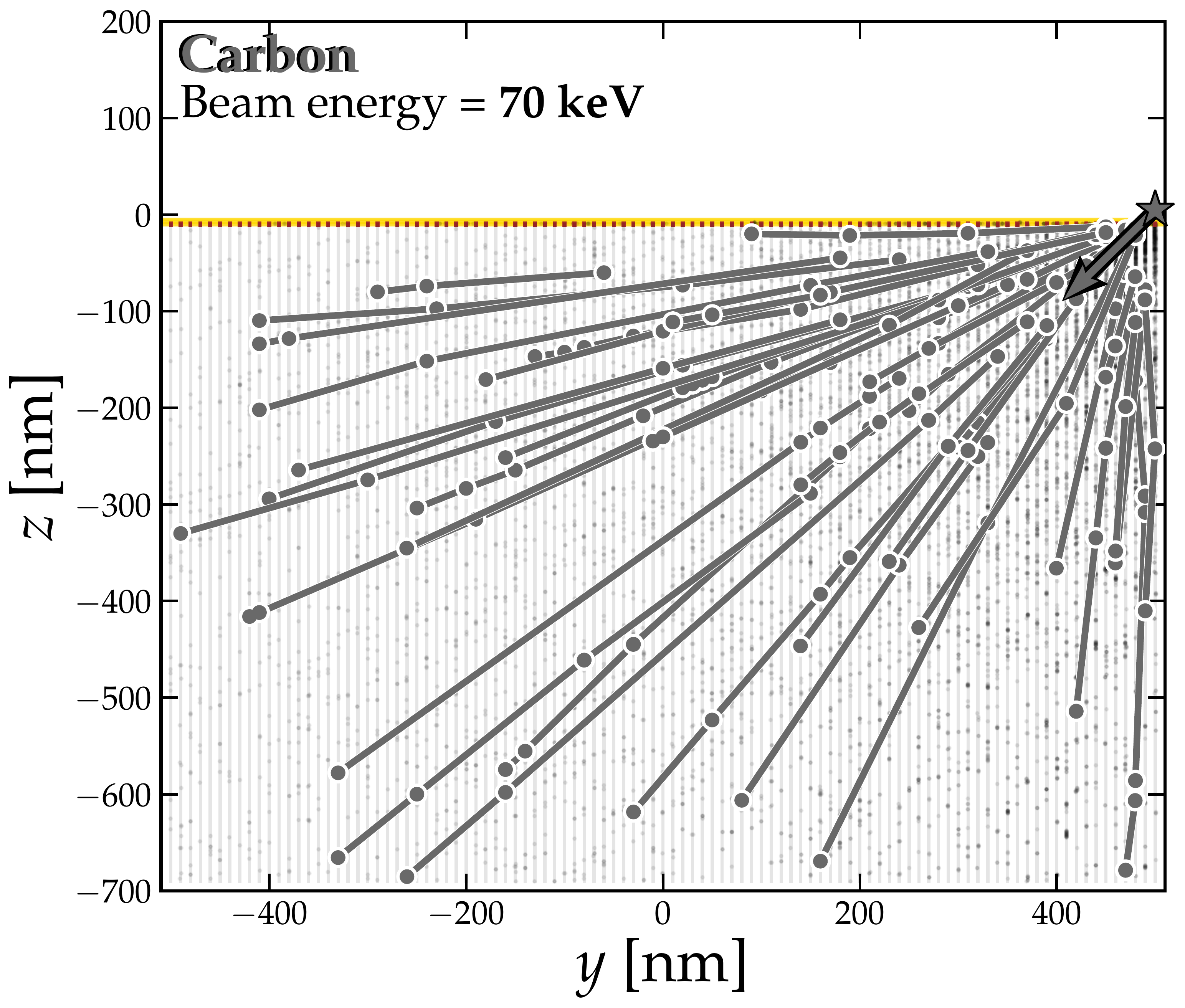} 
    \includegraphics[width=0.49\textwidth]{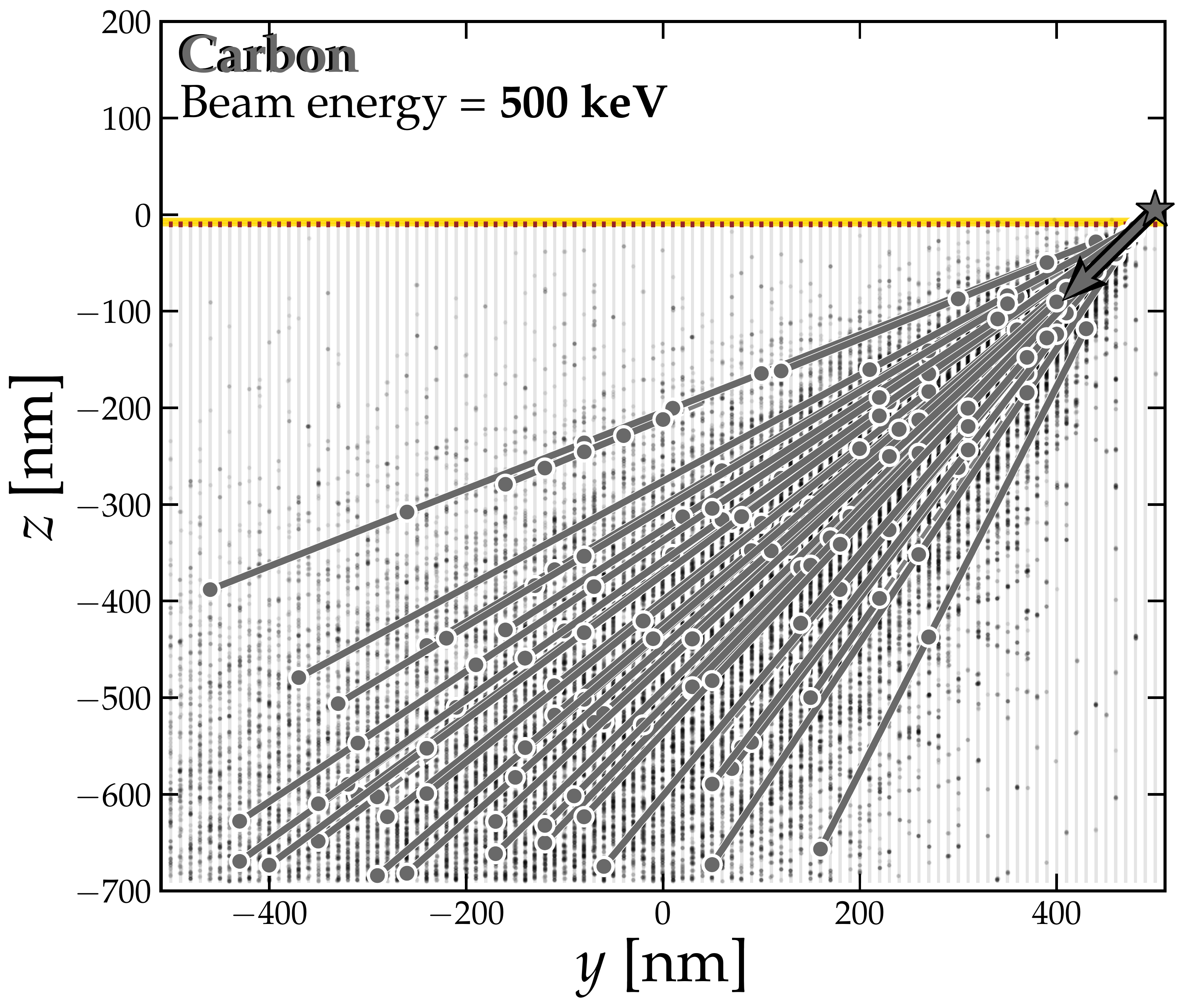} 
    \caption{\label{fig:trackexamples} Distribution of DNA breakage sites for 30 and 500 keV protons (top row), and 70 and 500 keV carbon nuclei (bottom row) incident on the holder (gold layer at $z=0$) at an angle of 45$^\circ$. The faint vertical gray lines represent DNA strands. Green and grey thick lines represent a random sample (out of a total of $10^4$ primary particles per beam) of 40 proton and carbon primary tracks, respectively, determined as `measurable' in that they break a sequence of more than three unique strands. Corresponding thinner dashed lines (most evident for 500 keV protons) denote secondary and tertiary particles which are also energetic enough to break strands. The positions of every counted breakage site (i.e. interactions above a cutoff of 10 eV) in the set of $10^4$ histories is represented by the black points.}
\end{figure*}

\begin{figure*}[t]
    \includegraphics[width=0.33\textwidth]{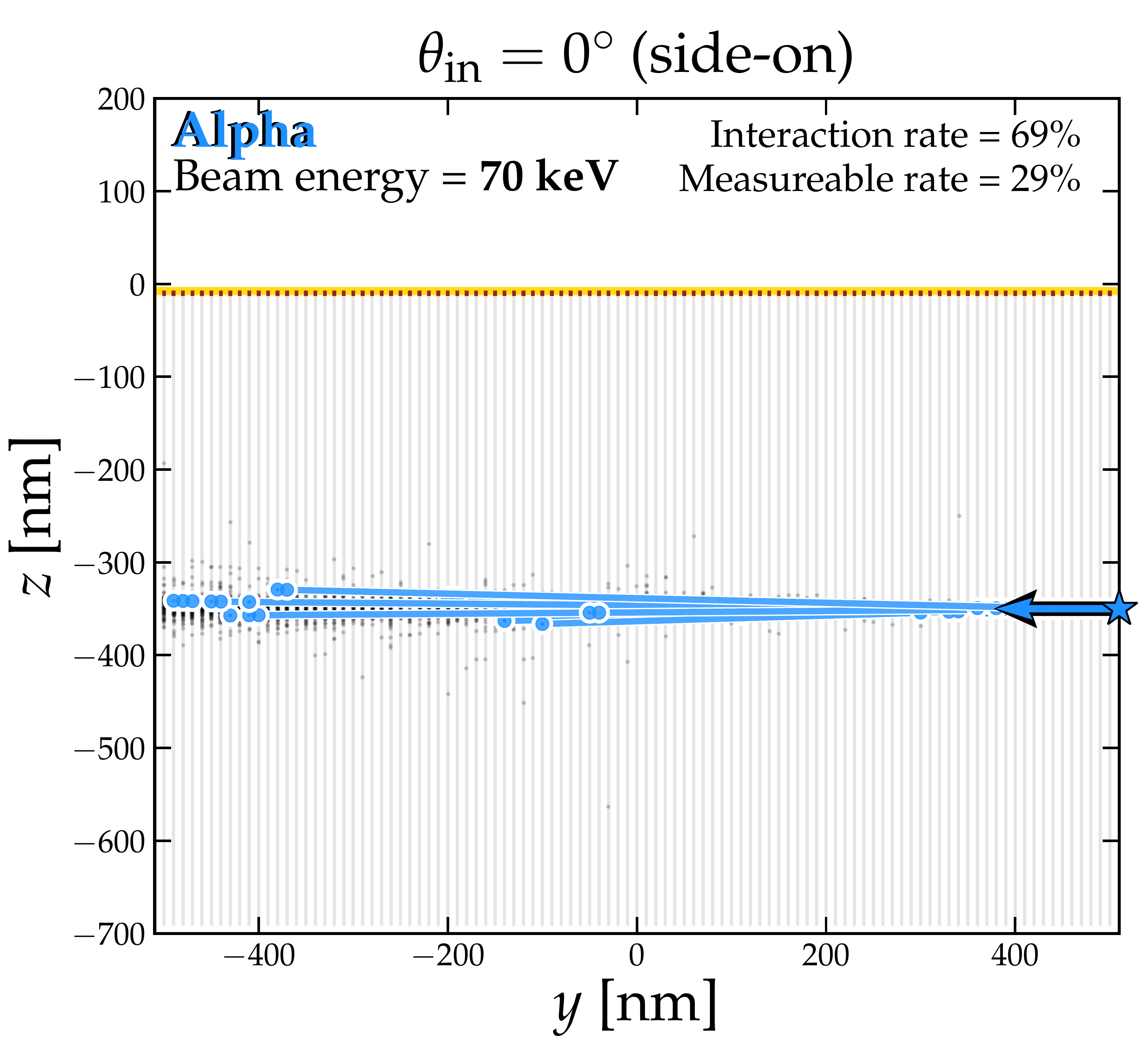} 
    \includegraphics[width=0.33\textwidth]{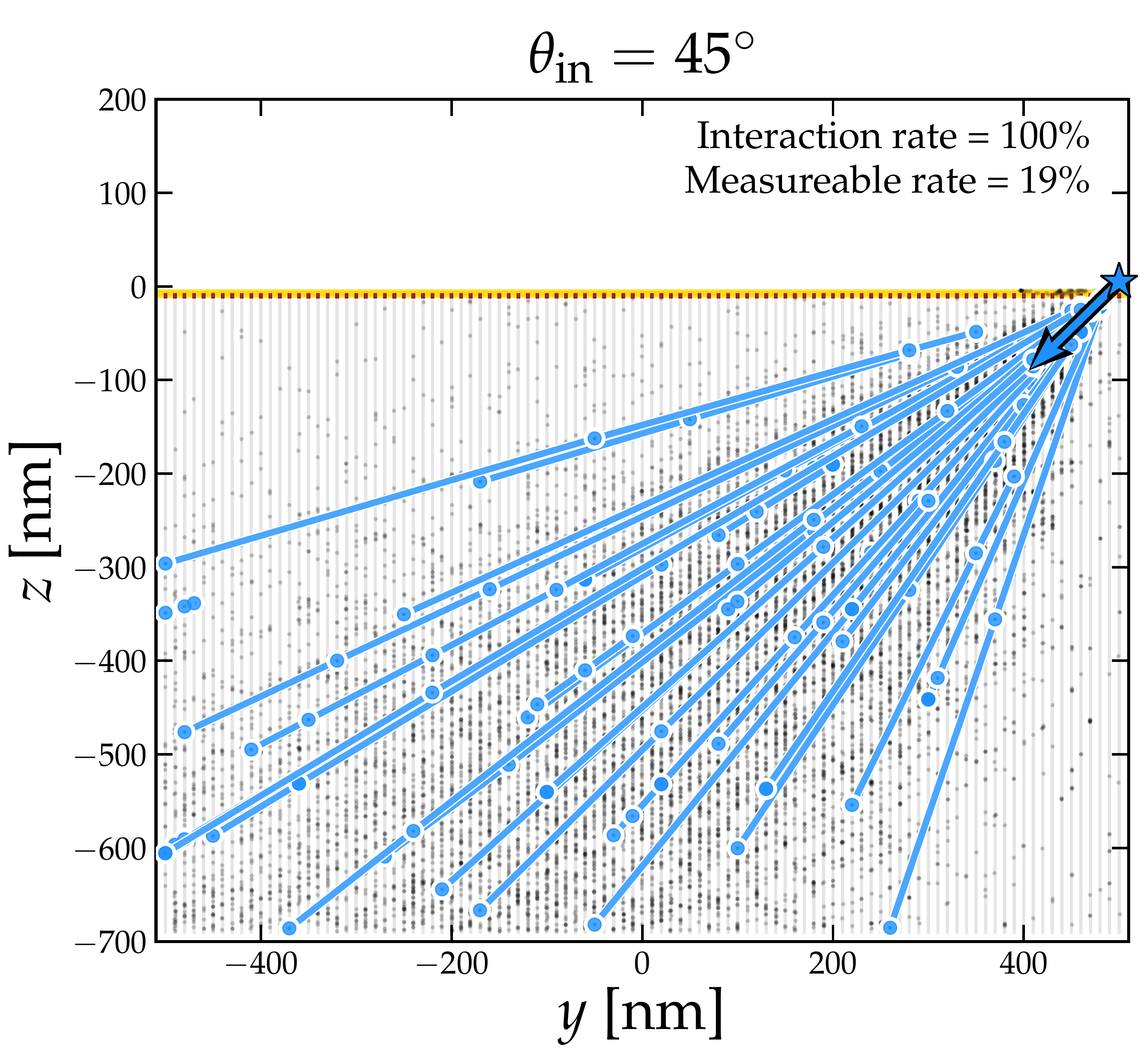} 
    \includegraphics[width=0.33\textwidth]{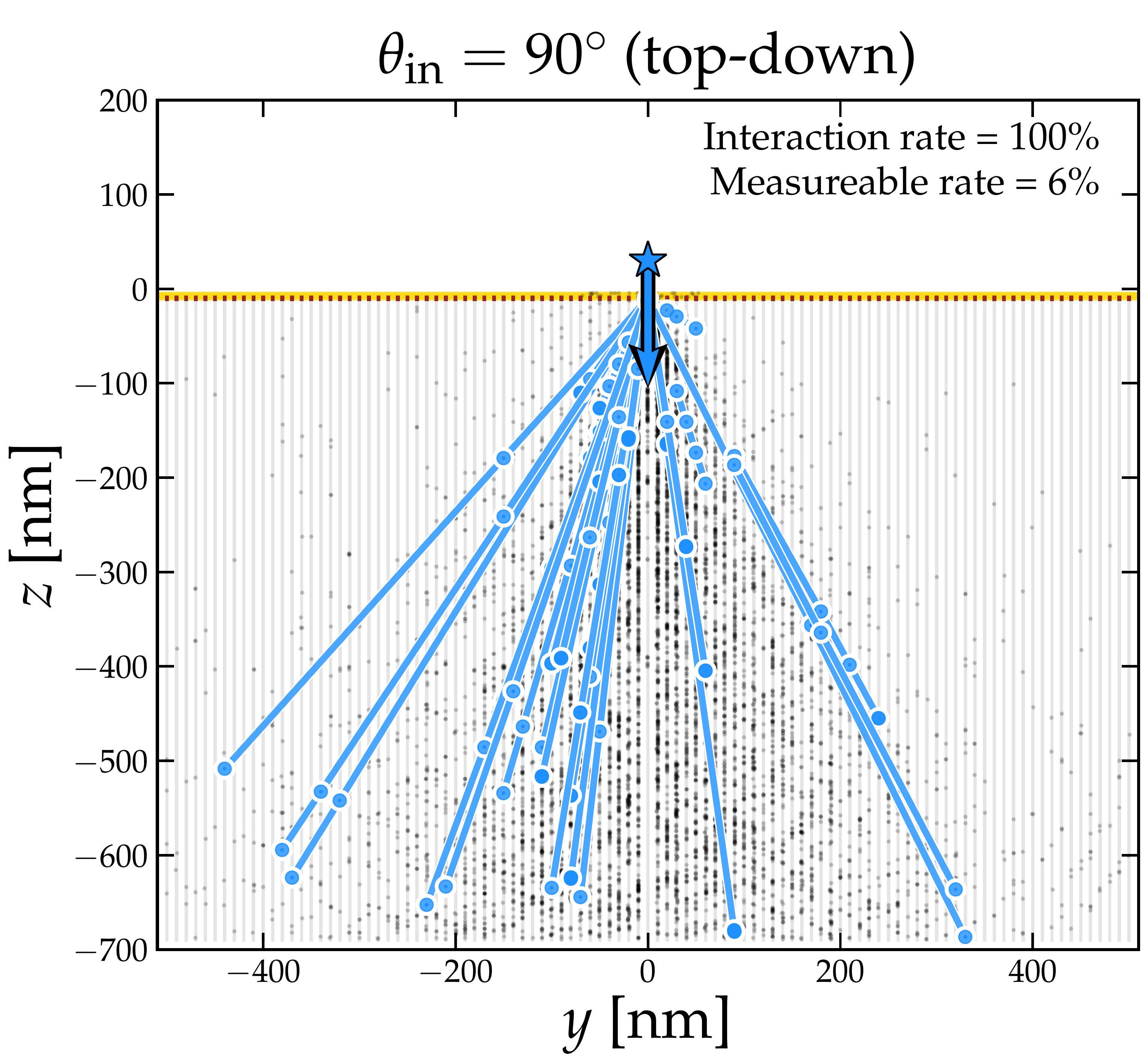} 
    \includegraphics[width=0.33\textwidth]{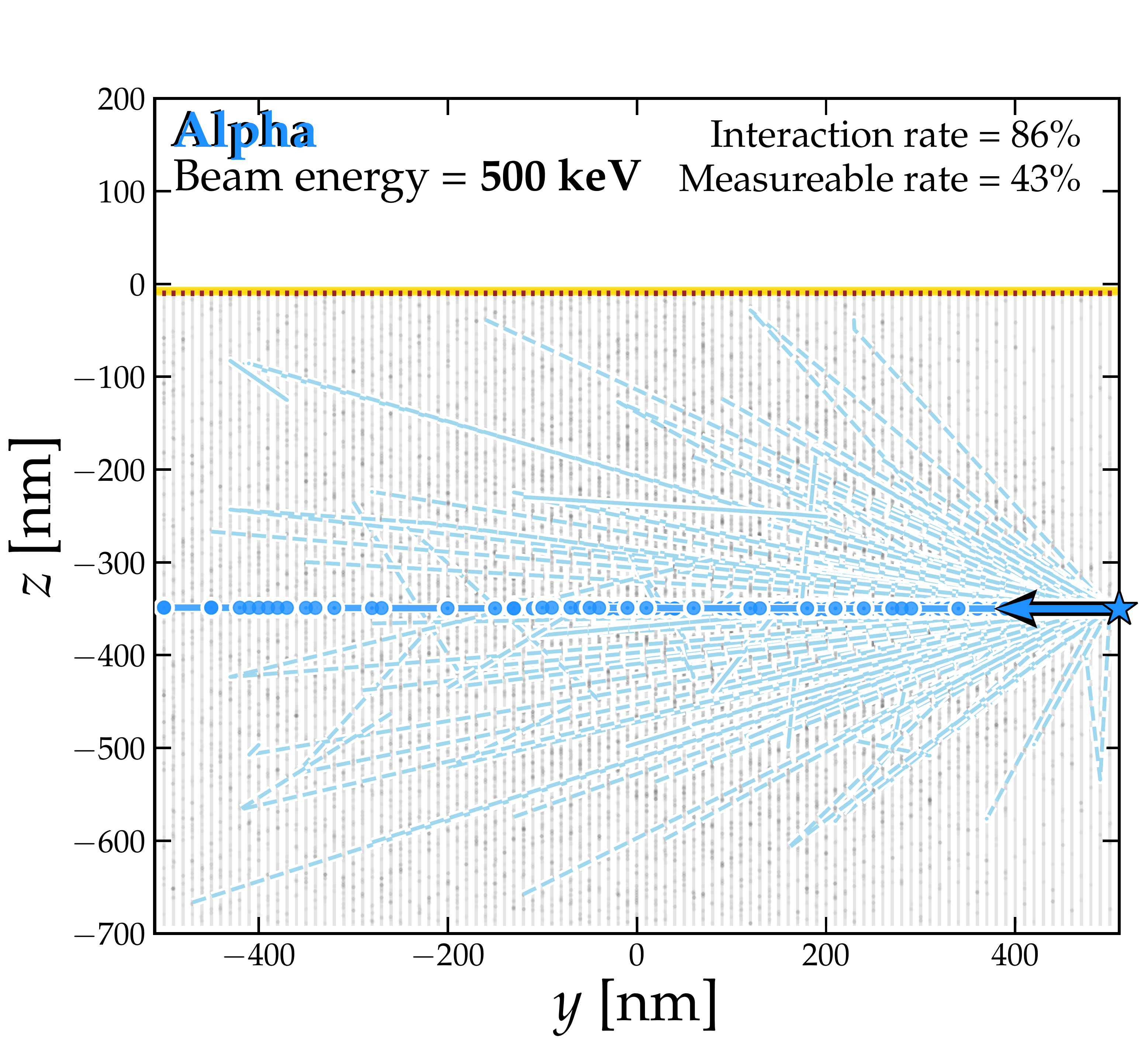} 
    \includegraphics[width=0.33\textwidth]{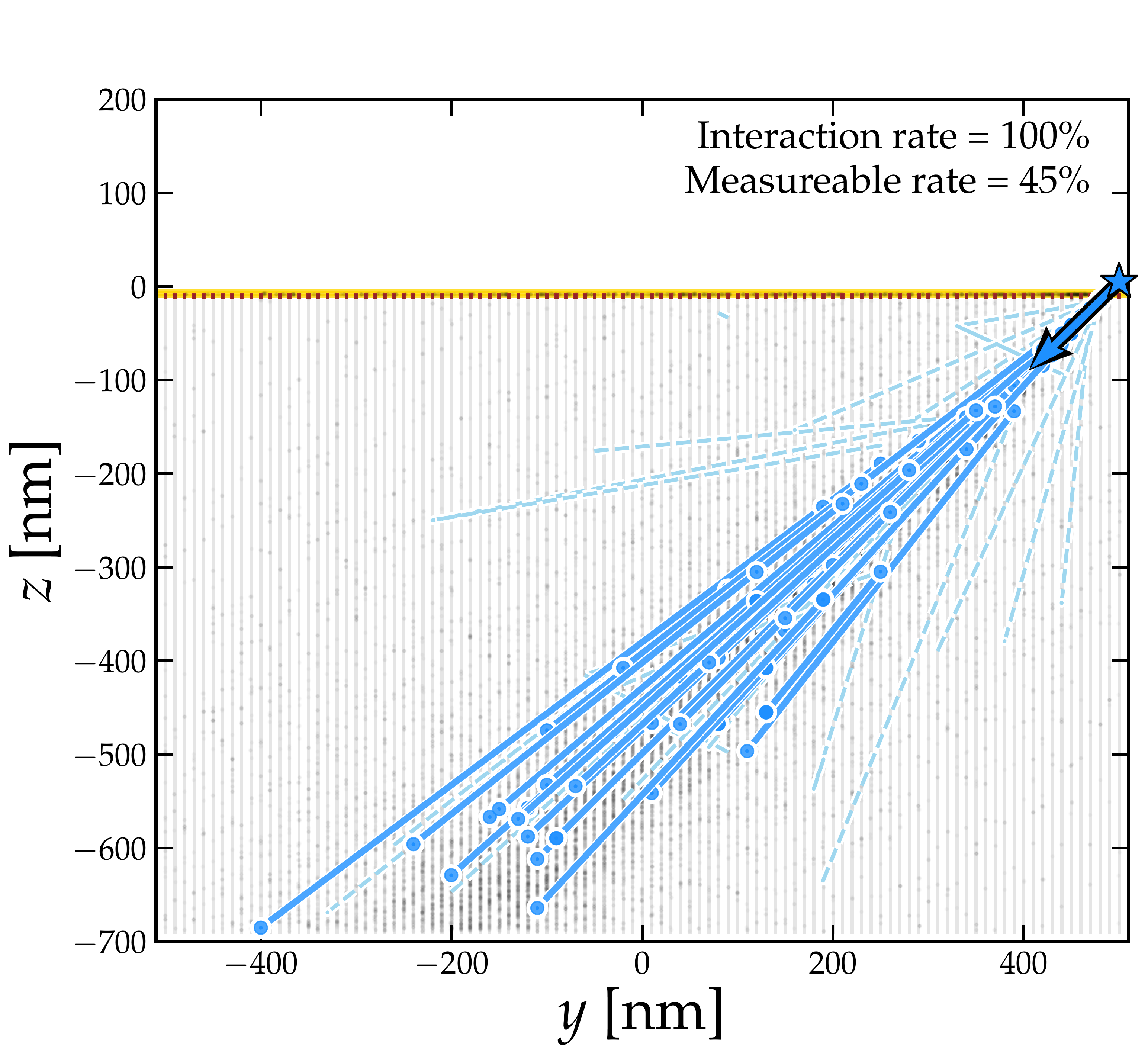} 
    \includegraphics[width=0.33\textwidth]{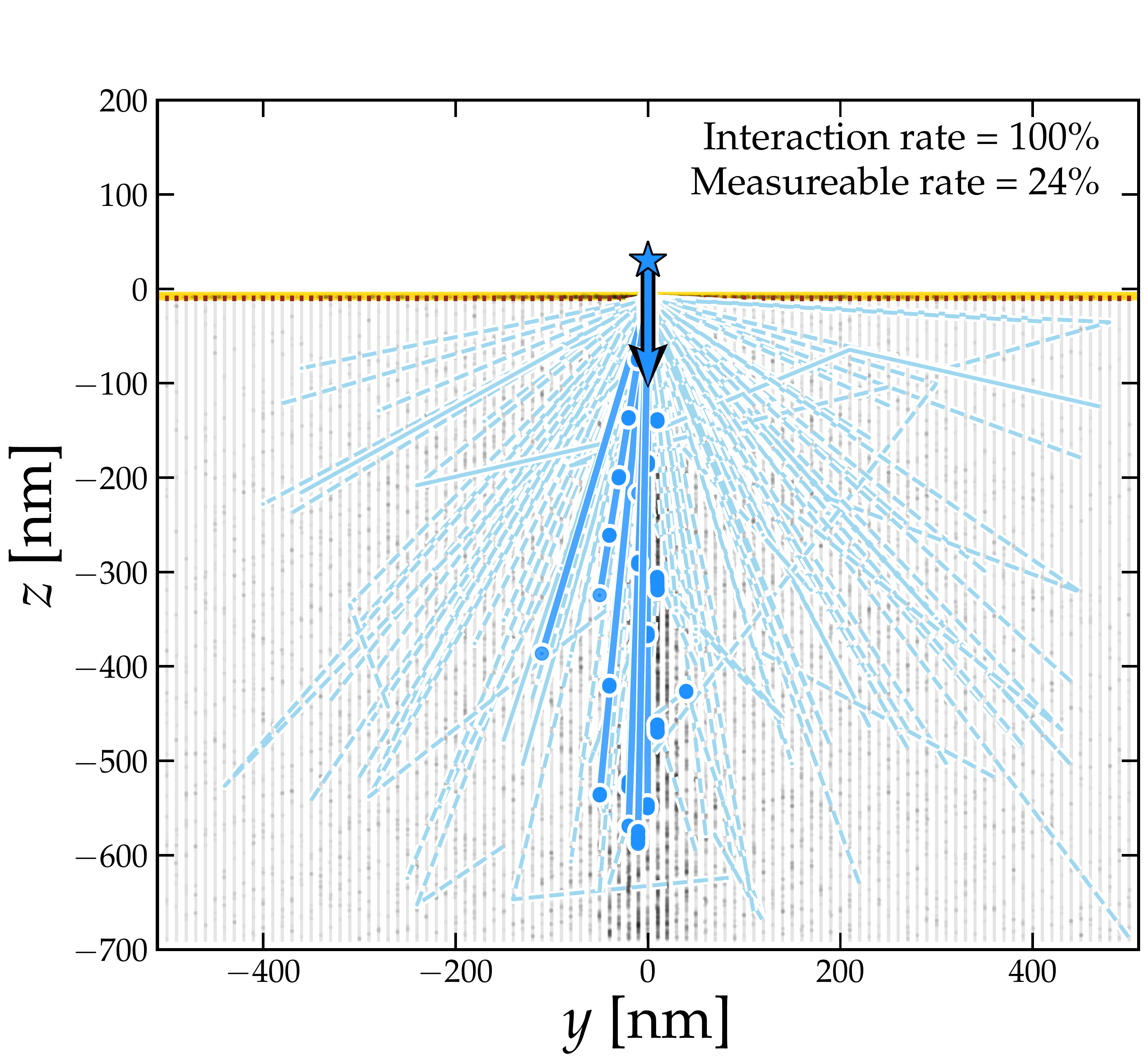} 
    \caption{\label{fig:trackexamples2_with_secondaries} Distribution of DNA breakage sites for alpha particles for two incident energies, 70 keV (top row) and 500 keV (bottom row), and for three different incident beam angles: $\theta_{\rm in} = 0^\circ$ (left), $\theta_{\rm in} = 45^\circ$ (middle), and $\theta_{\rm in} = 90^\circ$ (right). As in Fig.~\ref{fig:trackexamples},  breakage sites are denoted by filled circles and primary particle tracks by thick lines. Dashed lines (most evident for the 500 keV case) represent tracks of secondaries. A random sub-sample of 40 histories are shown out of a total of $10^4$ primary alphas. As in the previous figure the underlying black points show the distribution of all interaction sites. 
    }
\end{figure*}

\section{Simulation}\label{sec:simulation} 
Simulations were performed using the TOPAS-nBio radiobiology extension \cite{TOPASnBio} of TOPAS version 3.3 \cite{TOPAS}, which uses Geant4 version 10.6 \cite{Geant4_1,Geant4_2}. The interaction types simulated consist of multiple scatters, coupled transportation, and ionisation. The Geant4 physics models used include: low energy electromagnetic physics (Livermore model, g4em-lowep), stopping physics (g4stopping) and ion physics (g4ion). We did not use Geant4-DNA models \cite{Geant4-DNA} as the negligible effect on these simulations did not warrant the additional computational overhead. Instead, the simulations are more sensitive to the set energy threshold for DNA damage.

We fix a hard threshold of 10 eV below which interactions are not counted. If an interaction deposits more than 10 eV into a component volume of the strand then we consider the strand to be severed at that point. This threshold is slightly higher than the energy required to break DNA strands observed experimentally~\cite{DNAdamage-expt} and is consistent with the Geant4-DNA physics models \cite{Geant4-nanodos}. We show the probability of measuring an interaction defined in this way in Fig.~\ref{fig:PDNA}, for a range of incident particles. This figure demonstrated the ranges of energy and particle types that we are able to study.

\subsection{Setup}
Similarly to the way the strands are expected to be arranged inside the detector, in our simulation, we also assign each strand a unique ID. We assume that the only measurable information is the ID of the strand and the $z$ position at which it was cut. Primary and secondary tracks are distinguished in our processing of the data, and our visualisations. However, for our analysis of the tracks, we assume this distinguishing information is not known, unless it can be gleaned somehow from the track topology. Furthermore, if a single strand was broken in multiple places by a single track, we only assume we can reconstruct the position with the lowest $z$ (as discussed in Sec.~\ref{sec:eventrecon}). However, the number of instances of multiple breakage sites on a given strand for a single track is vanishingly small (even if the incident particle enters vertically). Put together, we end up with a minimum $x-y$ resolution given by $\Delta x=\Delta y = 10$~nm, and a $z$-resolution $\Delta z = 0.34$~nm. 

The primary particles from which all the Monte Carlo samples originate is referred to as the ``beam''. From the symmetry of our setup, we need only specify incoming particles with a vector defined by one angle, i.e.~$\cos{\theta_{\rm in}})$, where an incidence angle $\theta_{\rm in}=0$ corresponds to a particle travelling perpendicular to the DNA and parallel to the holder. We will generally show results for three scenarios: a beam incident from the top-down directly onto the holder $\theta_{\rm in} = 90^\circ$, a beam incident onto the holder at an angle $\theta_{\rm in} = 45^\circ$, and a beam incident directly side-on to the strands with $\theta_{\rm in} = 0^\circ$. Unless otherwise stated, the incident particles originate outside the detector.

\subsection{Events}
We perform simulations across a wide range of input parameters describing the beam: varying beam energies between $E_{\rm in} = 100$~eV to 10 GeV; $\theta_{\rm in}$ between $0$ and $90^\circ$; and beam particles including protons, alphas, heavy ions, electrons and photons. We do not explicitly simulate fluxes of particle generating elastic nuclear recoil events generated from the holder or the strands, e.g. neutrons or dark matter. Instead, to have more control over our results, we included gold and carbon nuclei as primary particle types; with the intent of interpreting their behaviour as if they were generated by some other earlier interaction, for example a heavy dark matter particle or cosmic ray generating a nuclear recoil. The directions of these nuclear recoils would be predicted by, say, the dark matter-nucleus scattering kinematics and would be what would need to be measured to confirm the anisotropy of a signal aligning with Cygnus as is done in a directional dark matter search. In fact, the DNA detector has an advantage over other approaches here because this \emph{initial} scattering angle is often very hard to measure in, say, a gas target. This is because the directions that are reconstructed are the full tracks which have undergone nuclear straggling, but the very small initial segment of the track that is required is typically smaller than the diffusion scale (see e.g. the discussion in Ref.~\cite{Vahsen:2021gnb}). In the DNA detector, the full straggled track that would pass through many layers would serve to find the initial event by coordinating several detector modules, but because we will show that direction information can be inferred at the single-unit level, this initial recoil angle can still be measured thanks to the lack of any diffusion. 

To first provide a general qualitative description of how these events appear in the detector, example visualisations are presented in Figs.~\ref{fig:trackexamples} and~\ref{fig:trackexamples2_with_secondaries}. Recall that events shown in these figures are thresholded to energy depositions above $10\,$eV. We show a more complete set of these visualisations across the full range of particle types, energies, and orientations in App.~\ref{app:additionalplots}.

Figure~\ref{fig:trackexamples} shows proton and carbon nuclei events and their subsequent breakage patterns for $E = 30\,$keV and $70\,$keV and $\theta_{\rm in} = 45^\circ$. A side-on ($y-z$) view is shown, with individual example tracks highlighted as lines and events indicated by the filled circles. These tracks are a random sample of the total for $10^4$ primary particles.
In these examples, we have drawn a line connecting the chronological sequence of breakage sites, however in our analysis we do not assume that the order of this sequence is known\footnote{Recall that it is the chronological order of the \emph{set} of tracks that our microfluidics stage is supposed to provide, not the chronological order of the break sites making up each individual track. Since the timescale over which a single particle track is deposited will be extremely short compared to the time it will take for the strand segments to fall down, we cannot assume that we will be able to determine the chronological order of the break sites in a single track.}. In the case of ions with relatively straight tracks, this lack of information is not problematic, however for lower momenta particles, such as low energy electrons, and very low energy ions, the scattering angles between each break site can become large, meaning a straight line fit will be much less well-correlated with the incident beam direction. 

Figure~\ref{fig:trackexamples2_with_secondaries} shows six examples of the breakage pattern for alpha particles, corresponding to three beam angles and two beam energies. The primary beam energy 500\,keV (bottom row) is above the threshold for which secondary electrons generated by interactions between the detector components and the beam have sufficient energy to break DNA strands on their own. These secondaries are marked as dashed lines, connecting them to the primary track that produced them. As before, we can detect each of these vertices, but the lines between them are only for display purposes.

\subsection{Tracks}
\begin{figure}[t]
    \includegraphics[width=0.49\textwidth]{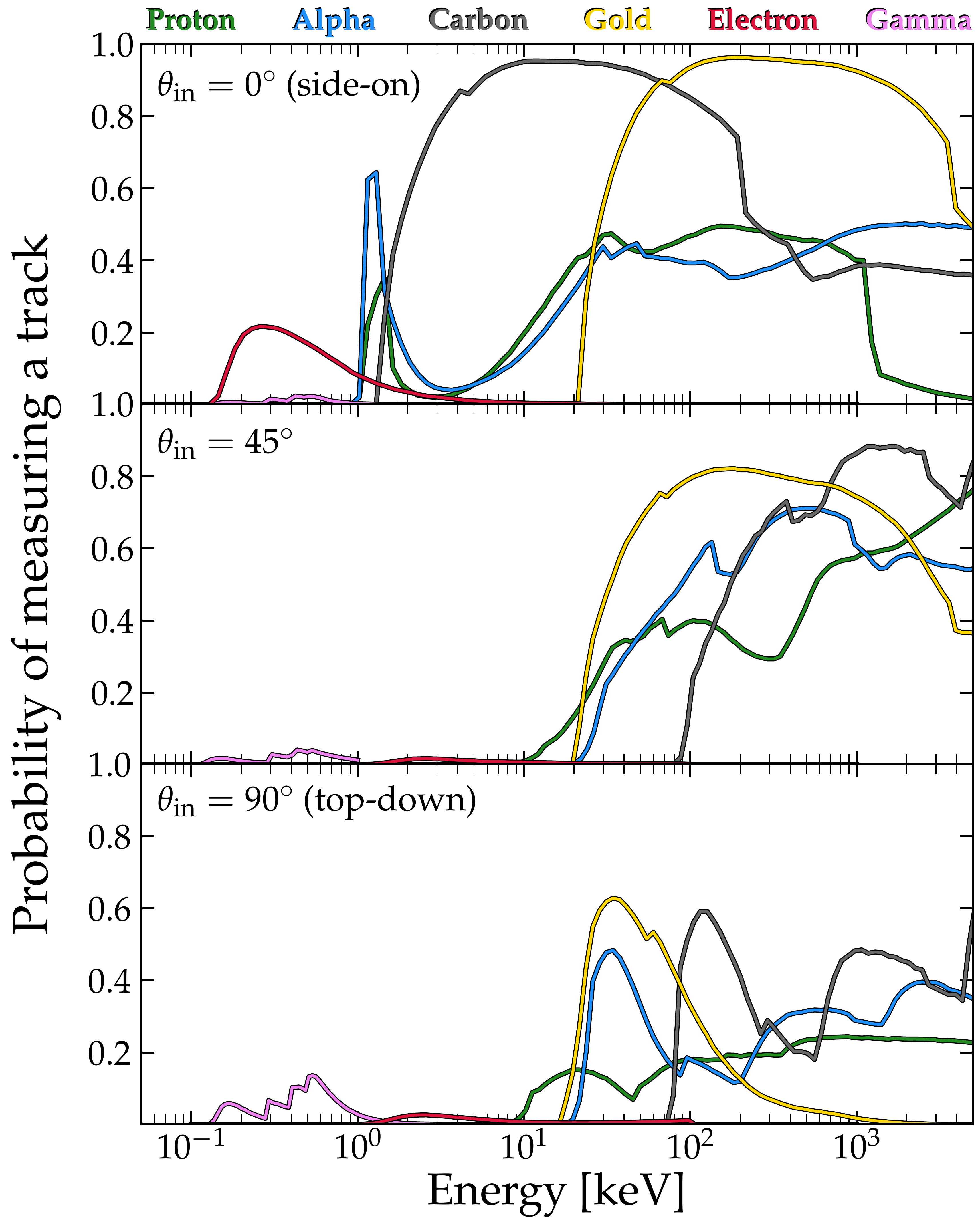}
    \caption{\label{fig:Ptrack} Probability of measuring a track for $\theta_{\rm in} = 0^\circ$ (top), $\theta_{\rm in} = 45^\circ$ (middle) and $\theta_{\rm in} = 90^\circ$ (bottom) beam angles as a function of primary beam energy. Colours correspond to different primary particle types, as indicated. A measurable track is defined to be one consisting of 3 or more events, each depositing more than 10 eV in a single DNA component volume.}
\end{figure}

Figures~\ref{fig:trackexamples} and  \ref{fig:trackexamples2_with_secondaries}, together with the additional figures shown in App.~\ref{app:additionalplots}, demonstrate  that typically, measurable ion tracks exhibit interaction sites that are coarsely spaced relative to the inter-strand spacing, i.e. there is typically a distance of $\mathcal{O}({\rm few}\times \Delta x)$ between each strand break. Occasionally, particularly for heavier ions that are more densely ionising, a sequence of neighbouring strands may be broken. This is evident for carbon ions in the lower panels of Fig.~\ref{fig:trackexamples}, particularly for the lower beam energy (70 keV). These simulations confirm that for the energies considered, primary ion tracks are straight and well-correlated with the incident beam direction, with increasing divergence from the incident beam direction with decreasing energy. 

For a track to be considered measurable, the bare minimum we could require is for three or more unique strands to be broken. To determine how likely these tracks are for different beam particles and energies, we calculate the probability $P_{\rm track}$ as the fraction of all primary particles simulated that result in a measurable track defined in this way.

This probability as a function of energy, particle type, and for three orientations, is shown in Fig.~\ref{fig:Ptrack}. We notice that the energy dependence of $P_{\rm track}$ is similar for the side-on and angled beams (top and middle panels), both resembling the energy dependence of the particle-DNA interaction probability, as shown in Fig.~\ref{fig:PDNA}. However, the case for the top-down beam is slightly different. This is because for this latter case, the incoming beam has to scatter by a certain amount to register more than one strand break. If the particle travelled through the holder and continued straight down, it would only break a single strand. So the structures seen in the shape of $P_{\rm track}(E)$ are due to the energies needed to scatter the particle to a neighbouring strand, without overshooting it and escaping out of the bottom of the detector. 

Figure~\ref{fig:Ptrack} informs us that the heavier ions (carbon and gold nuclei) are much more likely generate tracks inside this small detector module, with only a small suppression in the limiting case that an incoming particle strikes the holder directly from above. In the case of incident electrons, the probability of measurable track production is significantly reduced because of the attenuation in the holder (corresponding to the middle and bottom panels). Even for side-on incidence ($\theta_{\rm in} = 0^\circ$), electron tracks are much less correlated with the incident beam direction due to their propensity for wide scattering angles (cf.~Fig.~\ref{fig:ExtraBreakage_sideon} in App.~\ref{app:additionalplots}), which reduces their overall capacity for measurable track production. Photons are absorbed, and only create tracks when secondary electrons are excited with enough energy to break strands. Photons with energies between 1 and 100 keV pass through the holder unobstructed. The highest event rate is observed for the top-down beam because in this case the photons pass through the largest detector mass --- around $\sim$700 nm of a single strand --- as opposed to a much smaller amount in the other two cases. This maximises the probability of generating secondary electrons, and explains why the probability of having three or more interaction sites is higher than in the other two panels.

\begin{figure}[t]
    \includegraphics[width=0.4\textwidth]{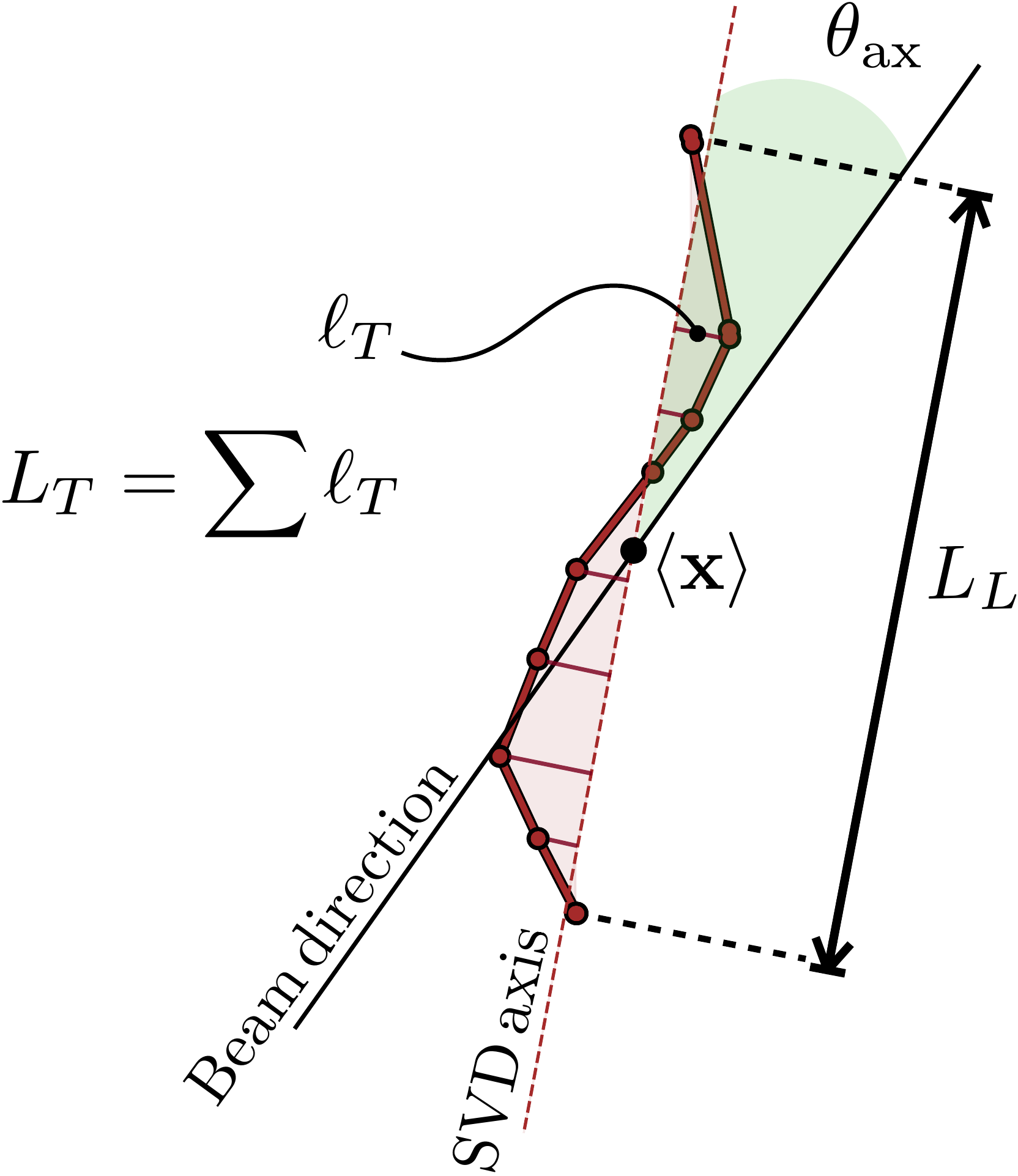}
    \caption{\label{fig:track} Diagram of geometrical quantities that can be extracted from a sequence of DNA strand break sites. We can use $L_L$ to parameterise the longitudinal length of the track and the $L_T$ to parameterise the transverse extent of the track. Since $L_L$ is typically larger than the detector, we can use $L_T$ to perform some comparisons between different particle types and energies. We also define $\theta_{\rm ax}$ to be the axial scattering angle.}
\end{figure}
In the following section, we compare the properties of the tracks produced by different particles. To calculate statistics about the simulated tracks, we need to fit them. We first obtain the primary axis of the track via a straight line fit using singular value decomposition (SVD). Then we use this distribution of interaction sites with respect to this vector to calculate various other characteristics, as shown in Fig.~\ref{fig:track}. For example, we can gain an estimate of the curvature of the track by looking at the size of the interaction site distribution in the direction transverse to the SVD axis (which we call $L_T$), and similarly we can also define a longitudinal track dimension $L_L$. We have checked the sensitivity of these quantities to the overall size of the detector, and find that $L_T$ is fairly insensitive. Therefore we use $L_T$ to compare the track distributions left by different particles. 

We can also calculate an axial scattering angle between the incoming beam direction $\hat{\mathbf{q}}_{\rm in}$ and the fit direction $\hat{\mathbf{q}}_{\rm fit}$,
\begin{equation}\label{eq:theta_ax}
    \theta_{\rm ax} = \cos^{-1}{(|\hat{\mathbf{q}}_{\rm in} \cdot \hat{\mathbf{q}}_{\rm fit}|)} \, .
\end{equation} 
We assume that no head/tail information will be present in the breakage pattern which is why we take the modulus of the dot product of the beam and track angles. If, for example, these two vectors were completely uncorrelated then the mean axial scattering angle would approach $57.3^\circ$ (i.e. 1 radian). Therefore a distribution of angles that is weighted towards smaller angles would be indicative of tracks that are correlated with the incoming particles. In contrast, distributions weighted towards larger angles (between $57.3 - 90^\circ$) would be anti-correlated on average. Therefore, only the cases where the mean value of $\theta_{\rm ax}$ approaches 1 radian correspond to no directional sensitivity at all.

\section{Results}\label{sec:results}
Assuming that the only information we can extract from the DNA detector are the $(x,y,z)$ positions of the broken strands, all the quantities shown in Fig.~\ref{fig:track} should be measurable. Using these metrics, we aim to determine the extent to which we can perform the usual tasks required of a particle detector, namely particle identification and energy reconstruction. Clearly, directional reconstruction is the major novelty of such a detector, so we will then evaluate the extent to which angular information can be measured as well. 

\subsection{Particle identification and energy reconstruction}
\begin{figure*}[t]
    \includegraphics[trim = 0mm 0mm 62mm 0mm,clip,height=0.7\textwidth]{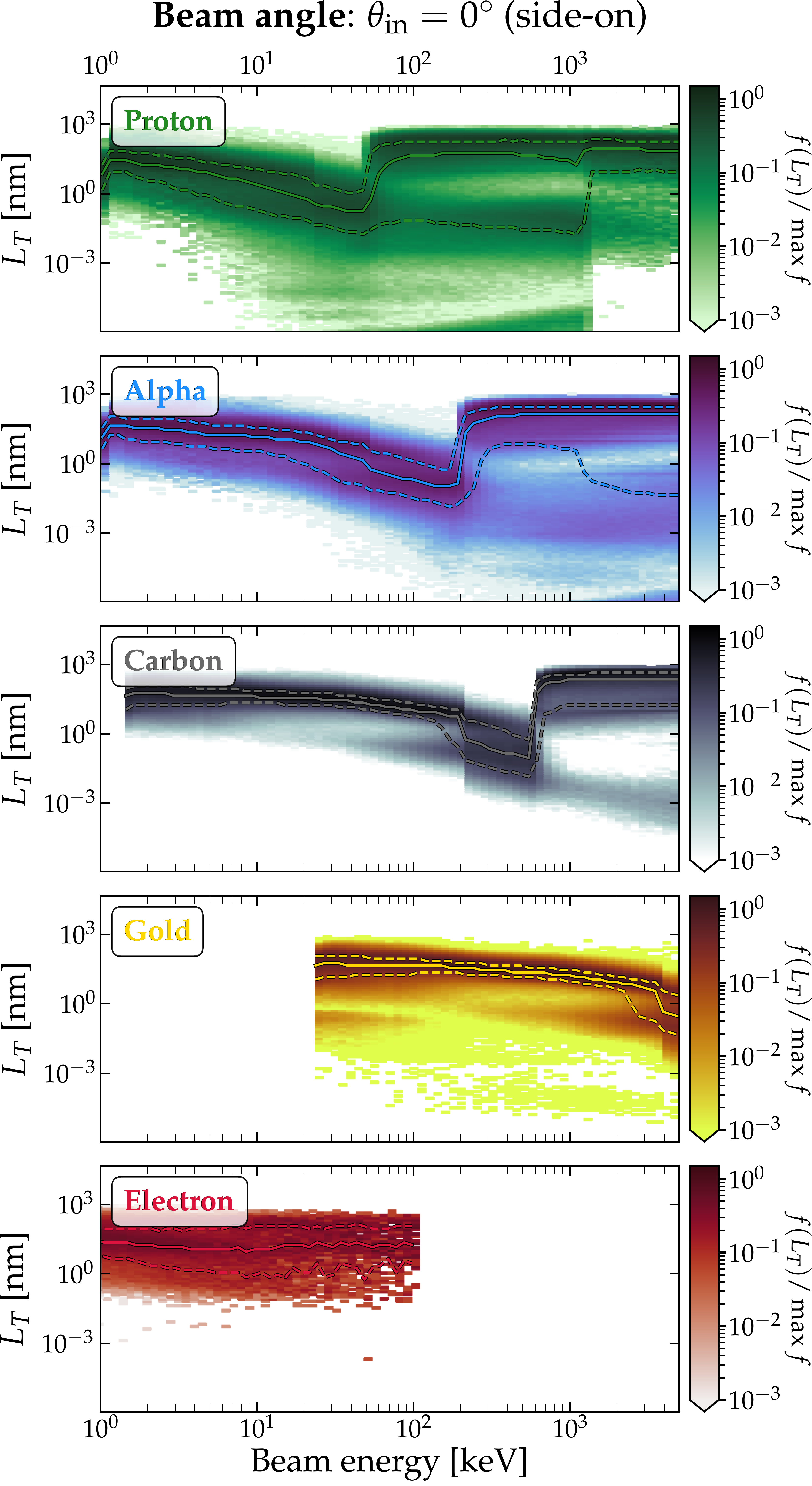}
    \includegraphics[trim = 20mm 0mm 62mm 0mm,clip,height=0.7\textwidth]{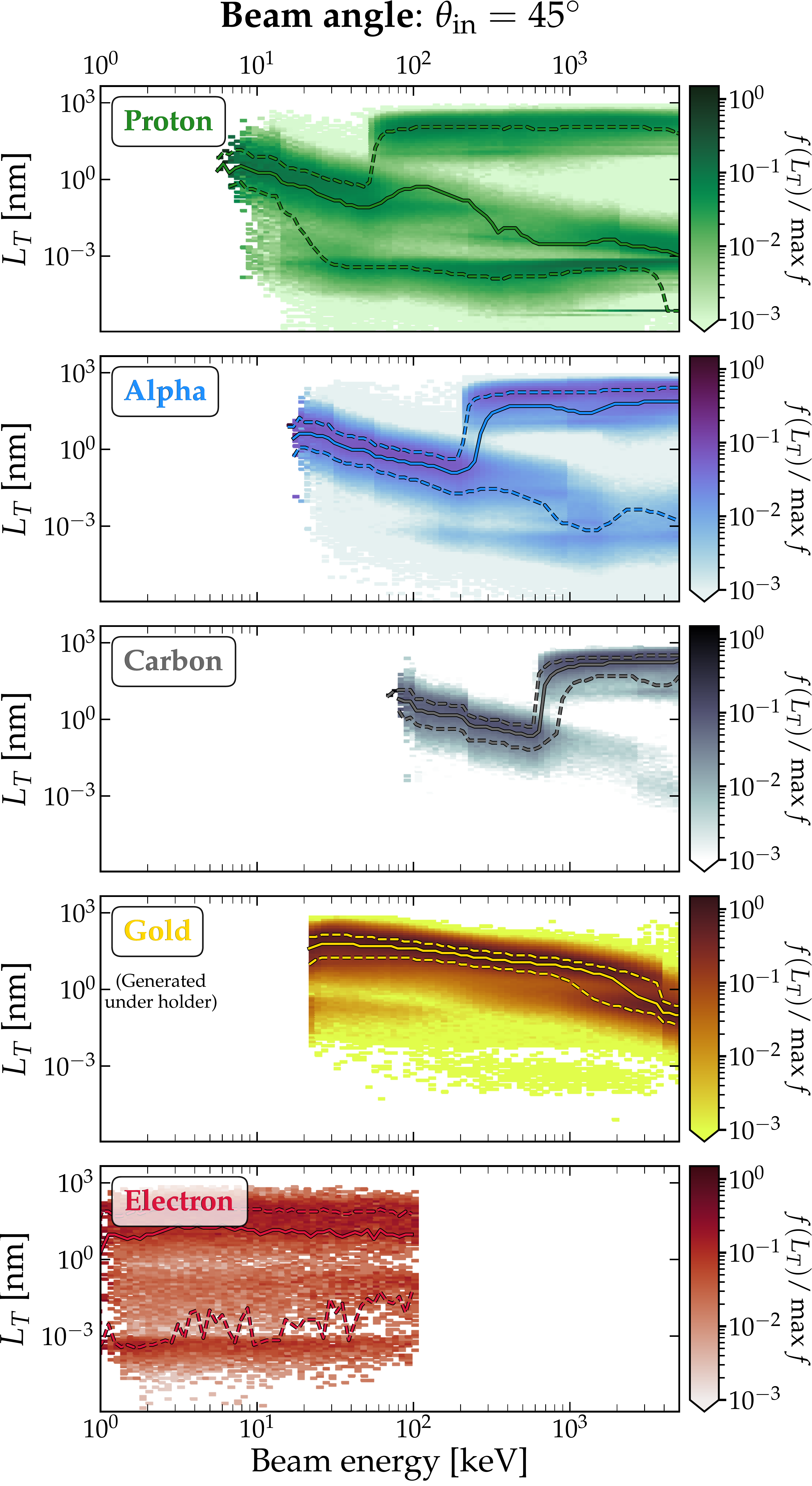}
    \includegraphics[trim = 20mm 0mm 0mm 0mm,clip,height=0.7\textwidth]{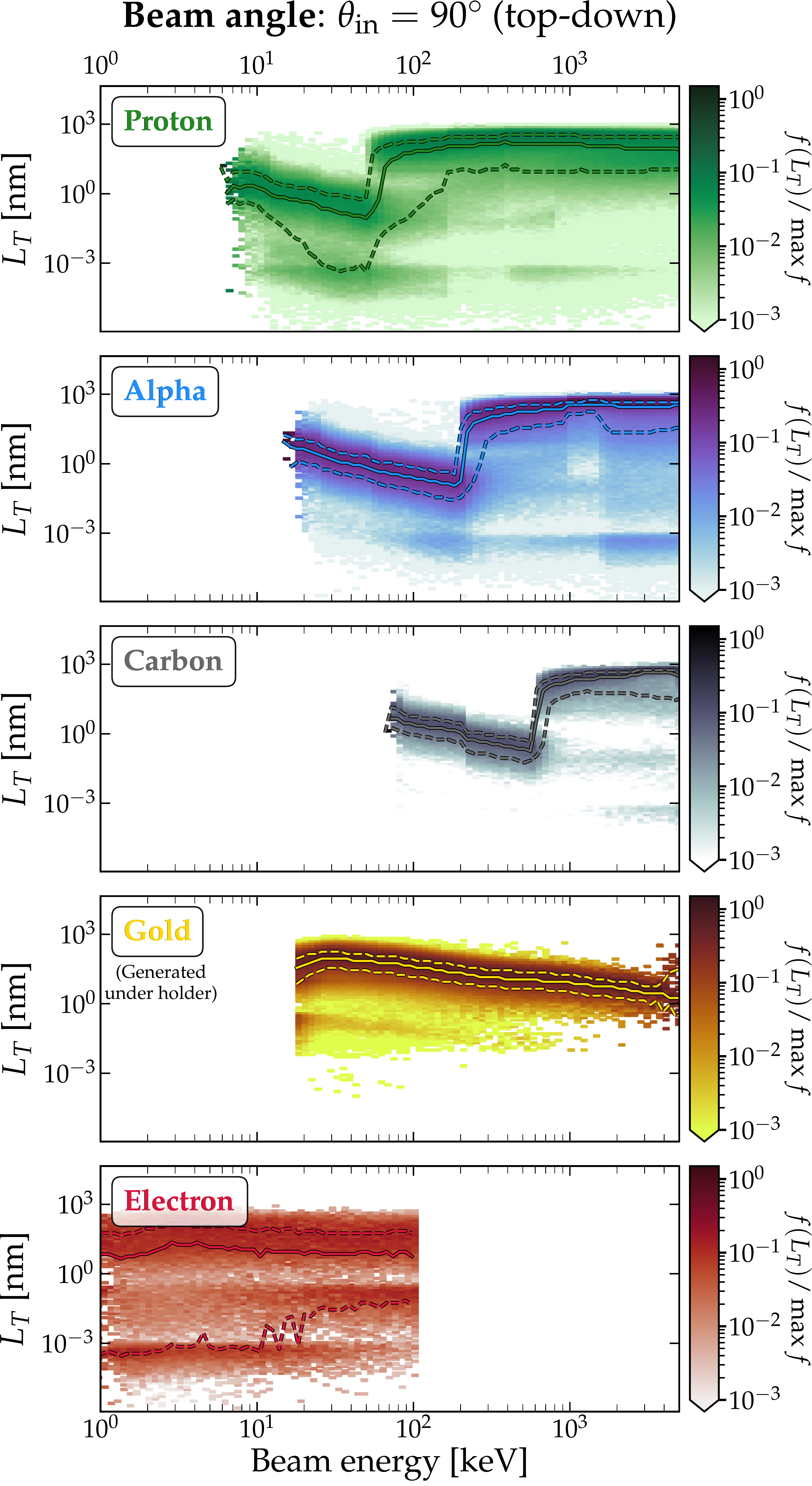}
    \caption{\label{fig:LT_dists} Distributions of transverse track dimension $f(L_T)$ as a function of incident beam energy for different incident beam angles. The colour-scale shows the value of the distribution, normalised by its maximum value at each energy. The four panels from top to bottom correspond to different incident particles: protons, alphas, carbon nuclei, gold nuclei, and electrons. Results are shown for gold nuclei in the $\theta_{\rm in} =0^\circ$ case for completeness and to help understand the differences between particles, despite the fact that gold nuclei recoiling from the gold holder with this incidence angle are not probable. Solid and dashed curves in each plot indicate the mean and 68\% containment of the distribution around the median.}
\end{figure*}
Since we only have position measurements to work with, any discrimination between impinging particle types and their energies will have to be done on a limited basis using only the topology of their tracks. As discussed above, the low density of the detector causes the tracks of primary particles to be generally straight over the scales we simulate here. However, for higher energies, the generation of secondary particles can be a source of noise (see the dashed lines in Fig.~\ref{fig:trackexamples2_with_secondaries}).

We present one approach that could be used to distinguish particle types in Fig.~\ref{fig:LT_dists}, where we show the probability distribution of the transverse track dimension $f(L_T)$ as a function of beam energy, for five different beam particles, and three different orientations (the same as shown in Fig.~\ref{fig:trackexamples2_with_secondaries}). The distinguishing features between the tracks left by the different particles shown here are primarily related to the momentum of the particle and the energy scale at which secondaries are produced with high enough energies to break strands. For instance, for low energy protons, alphas, carbon ions, and gold ions, the mean value of $L_T$ decreases with increasing energy because the tracks are straighter when the particle has a higher momentum. However, at a particular energy---e.g. at around 50 keV for protons incoming from the side (top left panel)---a large increase in the typical value of $L_T$ is evident due to the production of secondary particles. We also notice the structure of the distributions, which are noticeably multimodal at higher energies, forming distinct bands in the plots of Fig.~\ref{fig:LT_dists}. These bands are also associated with the production of secondaries, however the structures here are due to the discrete nature of the possible interaction sites, which are only allowed at integer values of $\Delta x$. Therefore when a secondary particle is produced, it must be generated in a direction where it will eventually encounter a strand and not just pass through gaps in between the strands. Thus, some values of $L_T$ are less likely, because they correspond to distances away from the primary track axis that are in between rows of strands.

We emphasise that while we display the distributions of $L_T$ on a logarithmic scale. Values smaller than $\sim$ 0.5 nm are likely not measurable. Also the upper tails of the distributions cut off at around 1~$\upmu$m due to the finite size of the box. Fortunately, almost all of each distribution is contained well within 200 nm or so. Also note that we have shown the size of the distributions on a logarithmic colour scale too, which also emphasises additional structure. Some of this may be relatively rarely occurring, e.g.~0.001 times less likely than the most likely value, which is where we cut off the colour-scale. The majority of these distributions are concentrated around a small range of values, close to their mean (shown as a solid line). We enclose 68\% of the distribution within the two dashed lines.

We conclude from these distributions that some level of particle ID may be achievable, but it will be strongly correlated with energy. At the most minimal level, it should be possible to say for an individual track whether the particle that caused it had a high momentum or low momentum. Depending on the expected rate of different incoming particles, this may be useful information for performing some low-level discrimination, or at least for selecting some candidate events. 

As for the reconstruction of the particle's energy, this too will be challenging for a single one of these small detector units. As with particle ID, we can see that only a relatively small quantity of information about particle energy is stored in the breakage pattern on these small scales. Therefore if this information is to be reconstructed on an event-by-event basis, this must be done on larger scales in a detector comprised of many smaller units. In fact, such a technique for particle ID and energy reconstruction could be rather conventional: for instance, by studying the differences in the amplitude and profiles of the $\mathrm{d}E/\textrm{d}x$ along the recoil tracks.

\subsection{Directional tracking}
\begin{figure*}[t]
    \includegraphics[trim = 0mm 0mm 62mm 0mm,clip,height=0.7\textwidth]{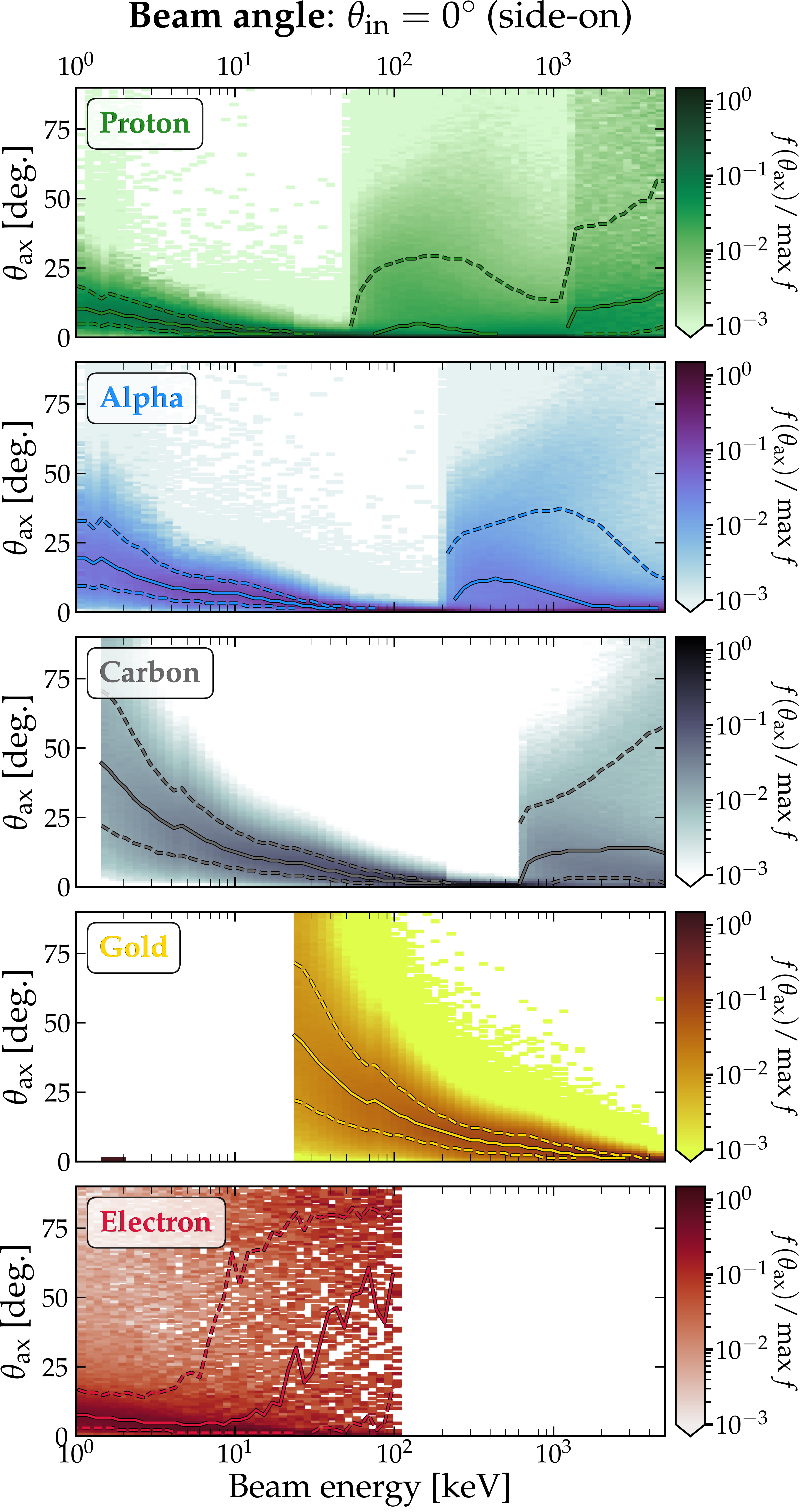}
    \includegraphics[trim = 20mm 0mm 62mm 0mm,clip,height=0.7\textwidth]{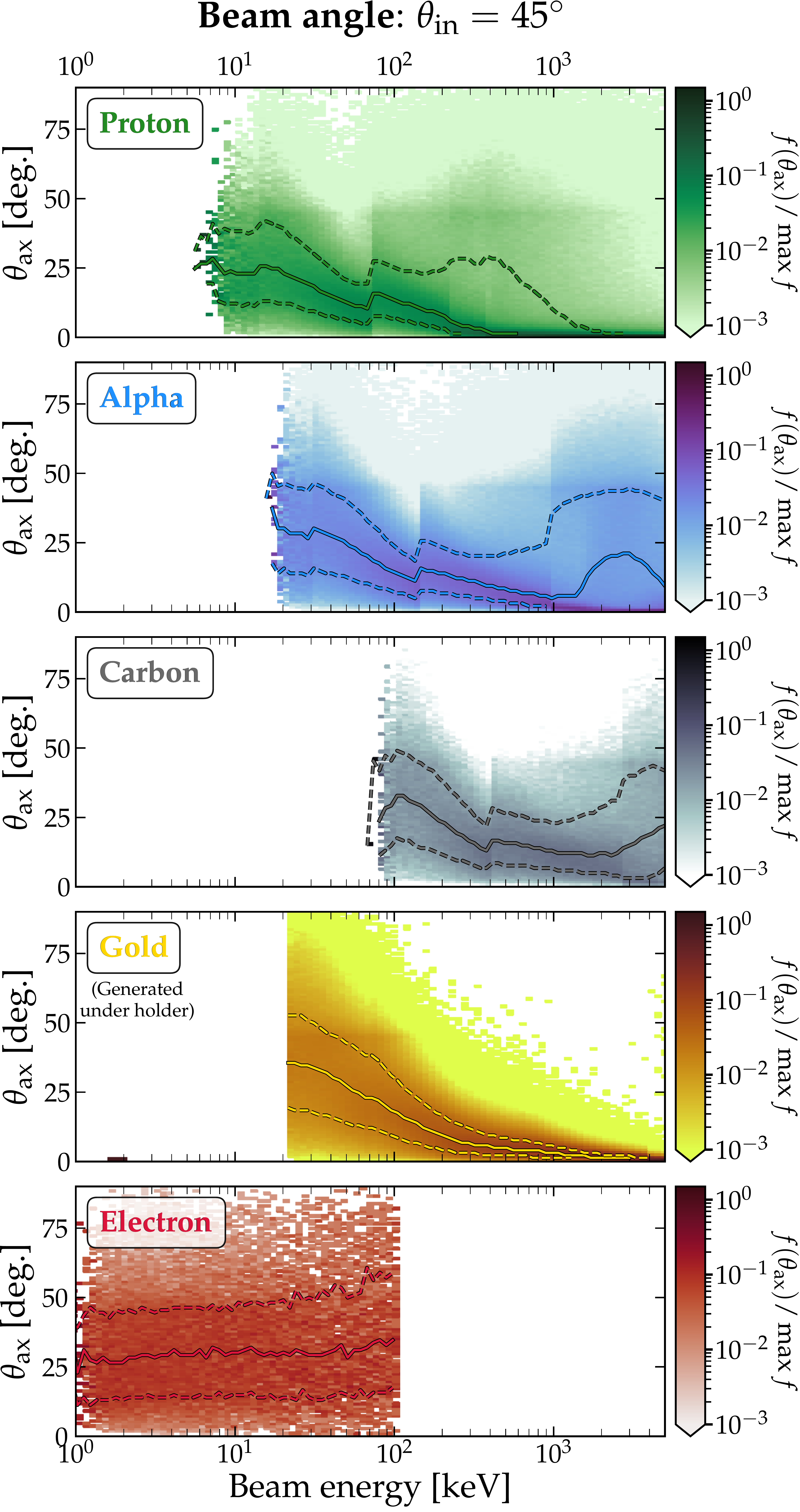}
    \includegraphics[trim = 20mm 0mm 0mm 0mm,clip,height=0.7\textwidth]{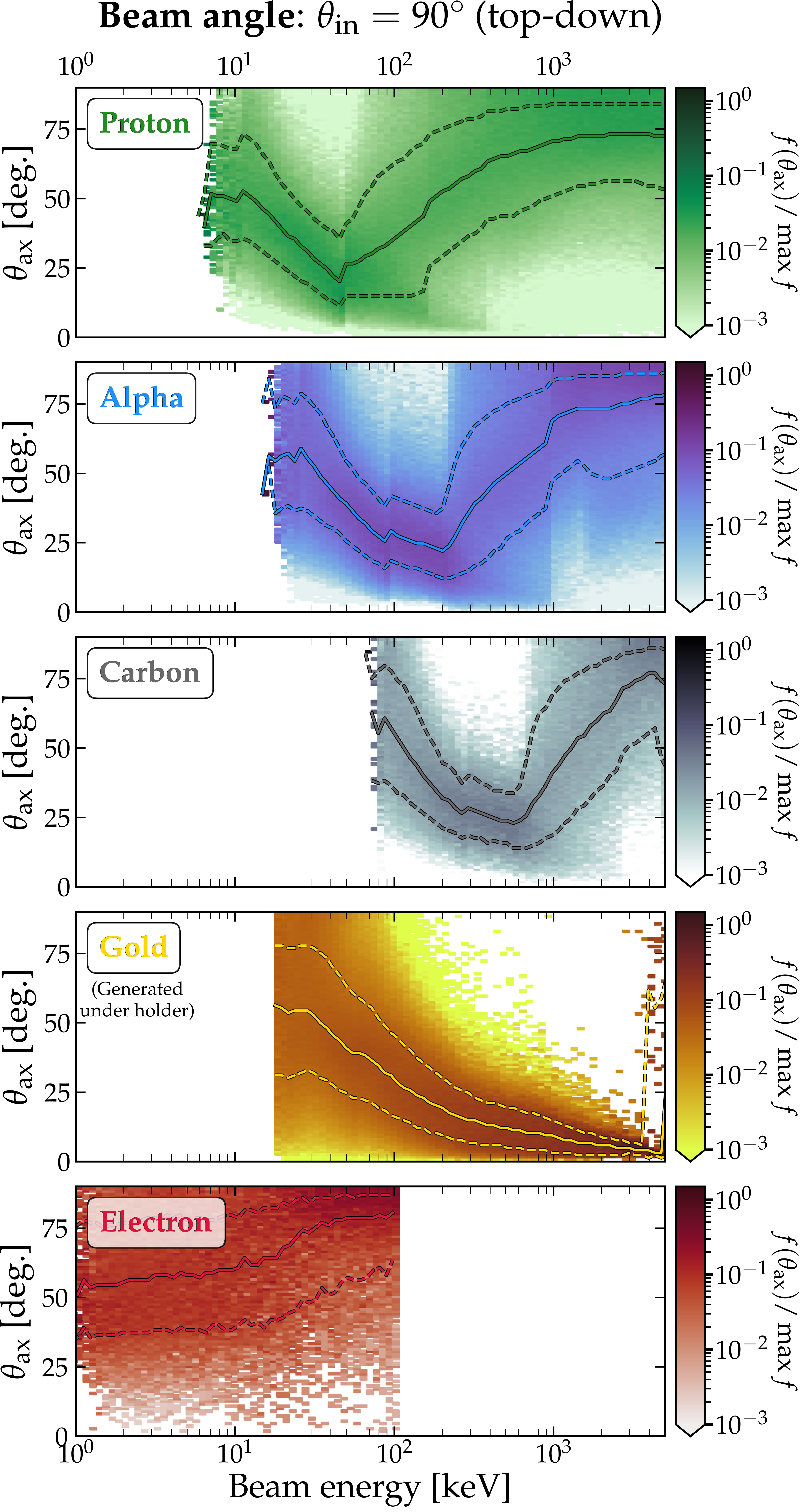}
    \caption{\label{fig:th_dists} As in Fig.~\ref{fig:LT_dists} but now for the probability distributions of the axial scattering angle $f(\theta_{\rm ax})$.}
\end{figure*}
We now finally evaluate the directional or angular performance of the detector based on the simulated tracks. Using the same SVD fit of the track direction, we parameterise the angular resolution associated with a particular particle type and energy by constructing a probability distribution of axial scattering angles $f(\theta_{\rm ax})$, where $\theta_{\rm ax}$ is measured with respect to the incident beam direction as in Eq.(\ref{eq:theta_ax}). 

Figure~\ref{fig:th_dists} shows the distribution of $\theta_{\rm ax}$ as a function of particle type, energy and for three different beam orientations. For $\theta_{\rm in}=0^\circ$ (side-on), the distributions for protons, alphas and carbon ions decrease with energy up to a threshold energy above which secondary electrons are produced with sufficient energy to break DNA strands. These secondary electrons scatter at wider angles away from the primary track, thus increasing the effective angular dispersion.

Overall, $\theta_{\rm ax}$ is generally well-correlated with the incident particle direction, and for some ranges of energy it is extremely well-correlated, with the full distribution of axial angles contained within 10 degrees. We also point out that even though the logarithmically coloured distributions extend to much wider angles for higher energy beams, the mean scattering angle (shown by the solid lines) is still relatively small for most particles for the $\theta_{\rm in}=0^\circ$ (side-on) and 45$^\circ$ cases. For $\theta_{\rm in}=90^\circ$, we note that some scattering angles are typically much wider than, and in some cases anti-correlated with, the beam direction (implied by angular distributions weighted towards angles greater than 57.3$^\circ$). However, this particular example is more sensitive to the bias we have introduced in the requirement that a track is formed from more than three interaction sites on \emph{different} strands. For this to occur for top-down beams, the particle has to scatter at least a small amount to reach a different strand from the one it is incident upon.

\section{Directions for future experimental investigation}\label{sec:outlook}
Having shown that there is potentially interesting information, particularly good directional information, that can be gained from this kind of detector, we now want to highlight the outstanding challenges in realising this ambitious idea experimentally. The first stage to demonstrate the feasibility in a practical setting would be to perform preliminary experiments on smaller scale analogues of each stage of the detection process. For steps that require the precise placement of DNA strands we will also require some method of validating that the DNA has been arranged as desired. While various microscopy techniques are available, we highlight the nm-scale imaging well below the diffraction limit of light enabled by DNA-PAINT~\cite{DNA-PAINT} as a particularly interesting and flexible approach.

\subsection{PCR/readout}
Since we have discussed the basic feasibility of the concept, we may now proceed to a more detailed design study for a prototype experiment that can validate the results of this study and test various options for basic operation. With the additional design clarity provided by laboratory studies, it will then become more feasible to evaluate the potential cost of such a detector. Fast, cheap and portable PCR machines for instance are highly-sought after for point-of-care medical diagnostics. Examples of cm-scale pocket-size PCR machines are already on the market at the scale of \$10 per unit. Detection ideas based on biotechnology in general can benefit from a much higher level of economic and market support, when compared with many technologies used in experimental particle physics. So we suggest it is not unwarranted at this stage to imagine a potentially low-cost full-scale experiment, despite the high level of complexity involved.

\subsection{Testing DNA breakage in the lab}
Another uncertainty that must be addressed is the physics of DNA strand breakage by ionising radiation when arranged in the complex setup envisaged for this detector. This physics could also be tested directly in the lab. DNA origami---structures of hybridised DNA arranged into nanoscale shapes---have already been used as rudimentary radiometers to test UV damage on DNA~\cite{Fang2020}. Similar studies could be done with a much wider range of particle sources. Most interestingly, would be if we can employ high-intensity beams on arrays of DNA nanostructures to systematically parameterise the effects of common sources of background in rare-event searches.

\subsection{Detector construction and strand configuration}
Next, we must also decide whether to use single or double-stranded DNA, or whether to opt for RNA instead. While the choice of DNA over RNA should have no impact on the detection capabilities of this concept, the double-stranded molecules would likely be preferable due to increased stability and reduced cross-reactivity between strands. However, this would need to be contrasted against the lower cost of using RNA.

Another major practical consideration that should be investigated is the details of how the strands and their base content should be synthesised, arranged, and sequenced. We have framed this study around the assumption that the positions would be encoded in the precise sequence of base pairs, however this is not the only option. In many ways, the methods that we could employ to arrange the strand network will share features with synthetic DNA storage systems~\cite{Skinner2017,Church1628,Goldman2013,Tabatabaei2015,Grass2015,Yazdi2017,Dong2020}. These can be thought of like punch cards made up of DNA: devices in which information is stored in some way on synthetic DNA strands that can be later extracted and read. Compared with conventional hard drive technology, storing information in base pair content is considered impractical because it is slow and potentially error-prone. However, Ref.~\cite{Tabatabaei2020} describe a way to encode information not in the base content but via ``nicks'' in the sugar phosphate backbone, which can be created in an accurate and parallel way using specially chosen enzymes. Speed and accuracy are essential when comparing DNA storage with more traditional technologies like hard-drives, but for our initial detector concept this may not be so important. We should keep this possibility in mind if we wish to scale the detector up. In this context, the nick sites would essentially encode the exact $(x,y,z)$ positions in binary, meaning the arrangement and reconstruction of the strands is almost trivial. At a larger scale, a nicking approach could greatly improve the efficiency since it circumvents the bottleneck of initially synthesising the strands with a precise sequence of bases. So if, for example, individual modules of a larger detector needed to be replaced at a frequent rate, this could be a way to facilitate that.

\subsection{Stability}
The final requirement of this detector that we are yet to touch upon is stability. We have implicitly assumed throughout that the detector is actually able to achieve long-enough exposures to observe a sizeable event rate of particles under low-background conditions. We do not fix any required stability here, but follow-up experimental tests should at least be able to check if the detector is unstable to a degree that makes it unusable. There are several reasons why this may be a concern. Possibly the most crucial of these will be the requirement that we keep the DNA strands straight. The proposed technique of attaching magnetic beads to each strand to ensure the strands fall to the readout, should also act to keep the strands straight~\cite{BustamanteDNATension}. However, for the detector to be stable, the strands must be able to support themselves against this pull until they are severed.

Another issue regarding stability involves shielding. This detector concept is peculiar in the sense that the act of detection essentially destroys part of the detector. Ensuring that the detector is constructed quickly and in a low background environment will therefore be optimal. As long as the event rate per detector unit was kept low (e.g. if the detector was constructed from the $\upmu$m$^3$-scale units we simulate here), then there should be no issues. Assuming that any broken strands appearing before the official start of the experimental exposure, were flushed away and not counted. 

\subsection{Reconstruction noise}\label{sec:noise}
The final issue that should receive more dedicated consideration in future studies is the fundamental limit to the accuracy with which the $(x,y,z)$ positions of the strands can be reconstructed. This will ultimately limit the tracking and directional reconstruction within a single detector unit, and therefore will be important to understand for measuring, for example, the initial segments of recoil directions. There are several ways in which sources of noise could enter beyond what we have accounted for in our simulations. These enter at the level of the strands themselves and at the readout stage.

Firstly, in terms of the strands, while we have fixed the $(x,y)$ resolution, in order to maintain good directional sensitivity this resolution would need to be matched in the $z$-direction as well. At the largest, the accuracy with which the heights of each base pair should be known should be on the order of the typical spacing between interaction sites. As we have seen this is typically at 10--20 times the interstrand spacing, or equivalently an accuracy in the $z$-direction of about 10\% assuming the same $L_z$ and $\Delta x$ we have adopted here. There are several reasons why obtaining this may be challenging. Firstly, as mentioned several times previously, free DNA strands tend to curl up. The technique we suggest of attaching magnetic beads to facilitate the strands' collection, has been shown to hold double stranded DNA under enough tension to maintain straightness~\cite{BustamanteDNATension}. However there is still a chance for the helix structure to remain curled up by several turns even after being straightened by the magnetic pull. This would reduce the overall length of the strand and lead to an error when reconstructing the true $z$ position of each strand break.\footnote{Alternative designs we have discussed that would fasten the strands at both ends may suffer less from this effect, so the $z$-accuracy could be a consideration that works in the favour of this kind of design}. Similarly if the DNA is not completely dehydrated when it is transferred from solution to the detector medium (e.g. air, or vacuum) then this will also induce an error in the length of the strand. Different DNA conformations that form in different media vary in length and the extent to which strands change length after placed in the detector should be measured experimentally first. 

To quantify the importance of these issues, if we desire that the final strand length to be within 100 nm of its fully straightened length, then we could tolerate around 30 turns of the helix structure coiled in the wrong direction while still maintain good accuracy. Note also that this error is also one-sided---the strands will always be shorter in the detector rather than longer. Furthermore, even if a spatial error of this size was not achievable, the $\Delta x$ spacing we have chosen here is still arbitrary, and optimising a larger detector would have the freedom to vary $\Delta x$ and $L_z$ until this uncertainty was brought down to a suitable level. Another strategy for limiting this error that could come at the cost of stability would be to use single stranded DNA which would be easier to keep straight.

The second way in which the spatial reconstruction of the break sites could be error-prone is at the readout stage. We highlighted earlier a few DNA sequencing techniques to reconstruct the severed strands, however in practice we would not be able to rely on this readout process to provided the location of each break site perfectly. For instance, techniques such as Illumina sequencing that would break up the strands into multiple smaller segments, so having large sections of each strand be identical to one another could lead to confusion in which ones were actually broken. A simple resolution to this issue would be to make the length of strand short enough for them to be sequenced in their entirety. Alternatively we could simply encode more unique information about the $(x,y)$ positions in the base information chosen during the detector construction, or perhaps with the use of nicks in the backbone, as mentioned above. Other sequencing methods such as Nanopore that could sequence longer strands while keeping them intact would have their own challenges. In particular the possibility for insertion and deletion errors that would lead to essentially stochastic noise in the reconstructed bases and hence the length, these would need to be dealt with at the analysis level by comparison with the known initial sequences, however we estimate that under our baseline geometry simulated here, even a 10\% error in the reconstructed sequence could be tolerated.

\section{Conclusions}\label{sec:conclusions}
We have performed the first comprehensive simulations of the DNA-based detector concept outlined several years ago in Ref.~\cite{Drukier:2012hj}. While a large number of practical details remain to be studied and experimentally verified, we have shown here at the very least that the basic idea is feasible: DNA strand breakage patterns can be used to reconstruct axial information about incoming particle directions. 

The precise nm-scale recoil vertex position reconstruction is evidently the key advantage of this detector, but this comes at the cost of limited to no particle ID and energy reconstruction performance. On these small scales the differences in the track topologies is minor and highly correlated. So any convincing particle ID and energy reconstruction would have to be performed by using the tracks left in multiple smaller units analysed in parallel.

Additional advantages of such a detector compared to other directional technologies are, for instance, the complete circumvention of diffusion, straightforward fiducialization along all detector dimensions, and potentially a relatively low cost per unit mass, however this remains to be investigated. Furthermore, we have proposed here that the previous detector design be extended to include a strategy of real-time DNA strand transportation via microfluidics. This step is likely to be crucial if the device is used in conditions that generate many events per unit detector volume. This is because we typically find that the vertices of particle tracks are not along neighbouring strands, but are typically separated by several multiples of inter-strand spacings up to a factor of 10. This means that in order to prevent many overlapping recoil events from scrambling each other, some separation in time will be crucial. Moreover, a transportation system would also provide a way to retain precise and chronologically ordered event time information, making it more suitable to detect astrophysical sources of particles such as dark matter or neutrinos~\cite{OHare:2017rag}.

While only a limited quantity of particle ID information is stored in the nanoscale breakage patterns, we do find trends in the various types of track topology that correlate with the particle momentum. Heavy particles occasionally break sequences of neighbouring strands in a characteristic way as they slow down. Electrons on the other hand generally create tracks with no clear direction, meaning they would not be easily reconstructable and would likely be a source of noise. A veto on events without three or more co-linear breakage sites should be sufficient to discriminate electrons from other sources of particle. It it is likely that further discrimination between particle types will be possible for a larger detector, so this study will need to be conducted next to find out.

While the tracking of the expected keV-scale tracks is readily achievable with existing detectors, combining good tracking capabilities, sensitivity to low-energy recoils and, most importantly, scalability to high target masses are all notoriously difficult~\cite{Vahsen:2021gnb}. This is why the most sensitive dark matter experiments currently do not attempt to measure directionality and instead opt for massive ton-scale target masses and extraordinarily low backgrounds to isolate a dark matter signal. We have shown here that the DNA detector can achieve good directionality, since the axial angles of the simulated tracks are generally well-correlated with the incoming particle that caused them. This may eventually mean that this design could work as a dark matter detector, however we need not restrict ourselves to this option.

For instance, many detector technologies are already able to reconstruct the incoming directions of cosmic rays, however usually only at higher energies than our focus here. This is the case for instance in semiconductor detectors~\cite{Essig:2011nj,Crisler:2018gci,Aguilar-Arevalo:2015lvd}, such as DAMIC~\cite{Aguilar-Arevalo:2015lvd}, which can detect ionisation energies in silicon as low as $50$~eV, using $25 \upmu{\rm m} \times 25 \upmu{\rm m}$ readout pixels. The precision in $\mathrm{d}E/\textrm{d}x$ that this capability implies means that DAMIC is able to access directional information. However it is diffusion-limited for low-energy recoils whose physical track lengths are shorter than $15 \upmu{\rm m}$. Neutrino detectors are another example of experiments that gain directional information. IceCube~\cite{Aartsen:2014gkd} for instance, with its 2400~m long strings set 125~m apart has been particularly successful in reconstructing the directions of ultra-high-energy neutrinos coming from astrophysical sources. There is a neat coincidence in concepts, comparing IceCube to the detector that we envision here. The design of the DNA detector is a similar concept to IceCube only on a miniaturised scale, and targeting much lower energy interactions.

Therefore, with the many potential uses of this detector for particle physics, and with the fundamental challenges that this concept presents to bio- and nanotechnology, we believe that we have collected a compelling case for pursuing this idea further. We hope that this first explorative study will inspire imminent experimental investigation to resolve many of the outstanding issues we have laid out, and that future simulation studies can be better informed by the experimental realities involved.

\section*{Author contributions}
CAJO performed the simulations and analysis leading to the results presented here, and led the writing of the manuscript. JH, RJ, WK, VM, JN and KS assisted in the setting up and running of the simulations as part of short-term undergraduate projects. ZK provided setup files that facilitated in the initial creation of the simulations. AM provided a TOPAS-nbio extension file for the DNA detector. CB and SNG suggested the possibility to detect cosmic rays with this kind of detector and started this investigation.  CB, SNG, ZK also guided the overall scope of the paper, devised directions for future experimental validation, and helped write the manuscript. 

\section*{Acknowledgements}
We thank Giuseppina Simone, Katie Freese and George Church for collaboration during the early stages of this project, and Shelley Wickham for discussions. CAJO and CB and greatly appreciate support from the University of Sydney. CAJO is also supported by the Australian Research Council under the grant number DE220100225. SNG gratefully acknowledges TMS funding (BFS2017TMT01).

\bibliographystyle{bibi}
\bibliography{biblio}

\newpage

\onecolumngrid
\appendix
\section{Additional breakage plots}\label{app:additionalplots}
\begin{figure}[t]
    \includegraphics[trim = 0mm 15mm 0 0mm,clip,width=0.9\textwidth]{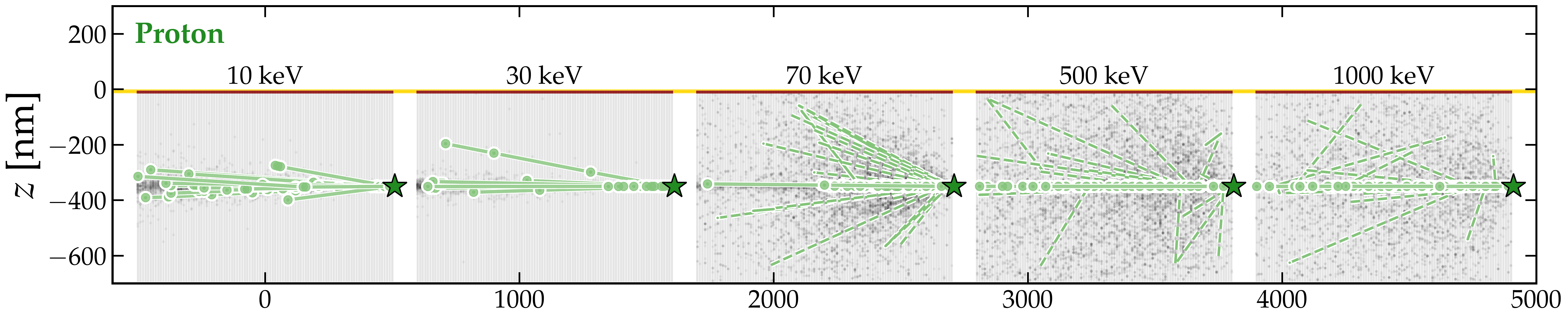}
    \includegraphics[trim = 0mm 15mm 0 0mm,clip,width=0.9\textwidth]{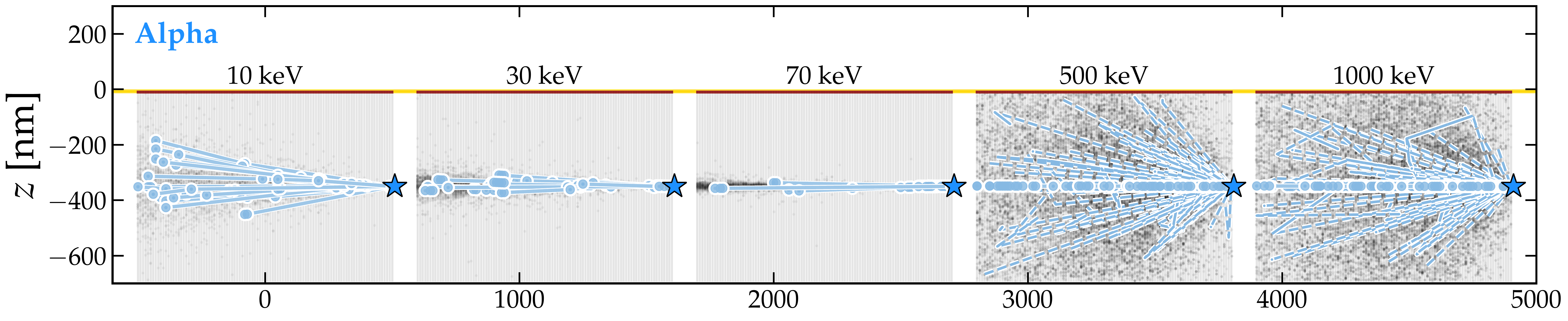}
    \includegraphics[trim = 0mm 15mm 0 0mm,clip,width=0.9\textwidth]{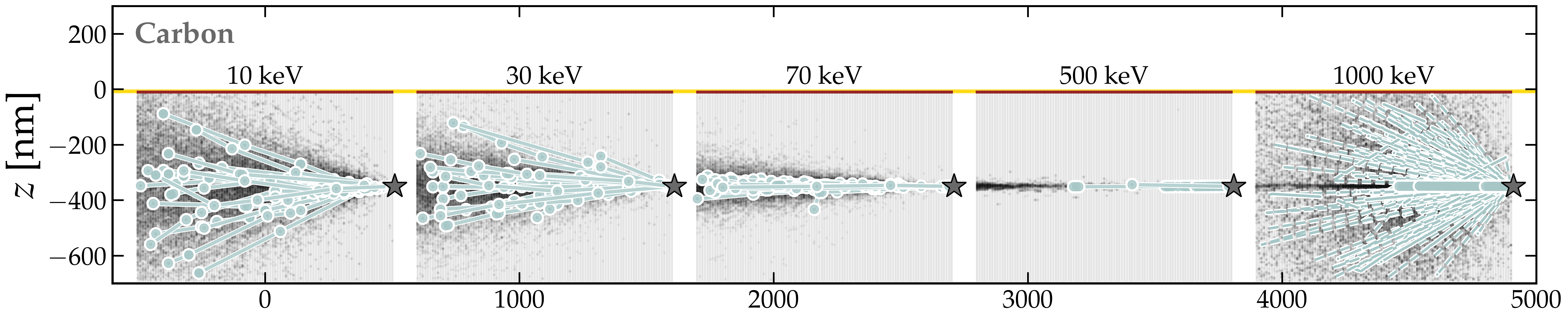}
    \includegraphics[trim = 0mm 15mm 0 0mm,clip,width=0.9\textwidth]{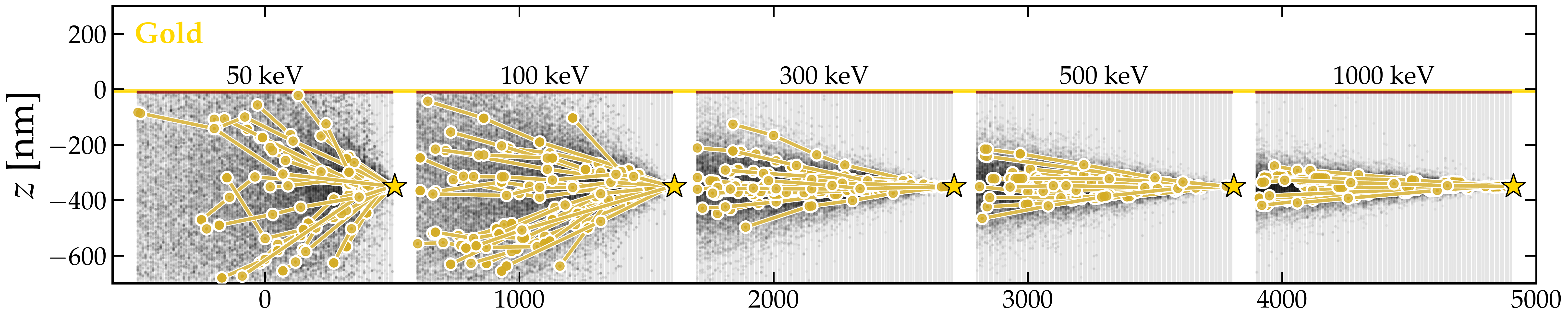}
    \includegraphics[trim = 0mm 15mm 0 0mm,clip,width=0.9\textwidth]{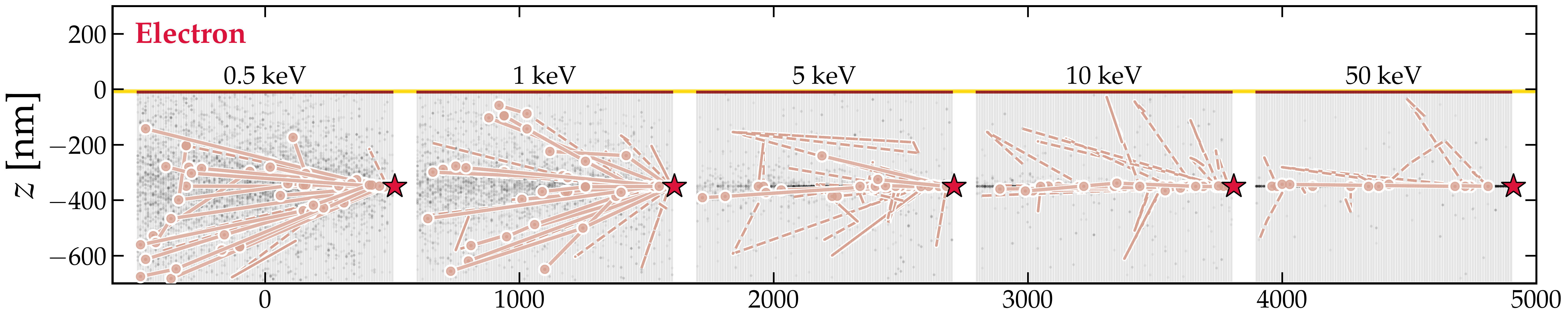}
    \includegraphics[width=0.9\textwidth]{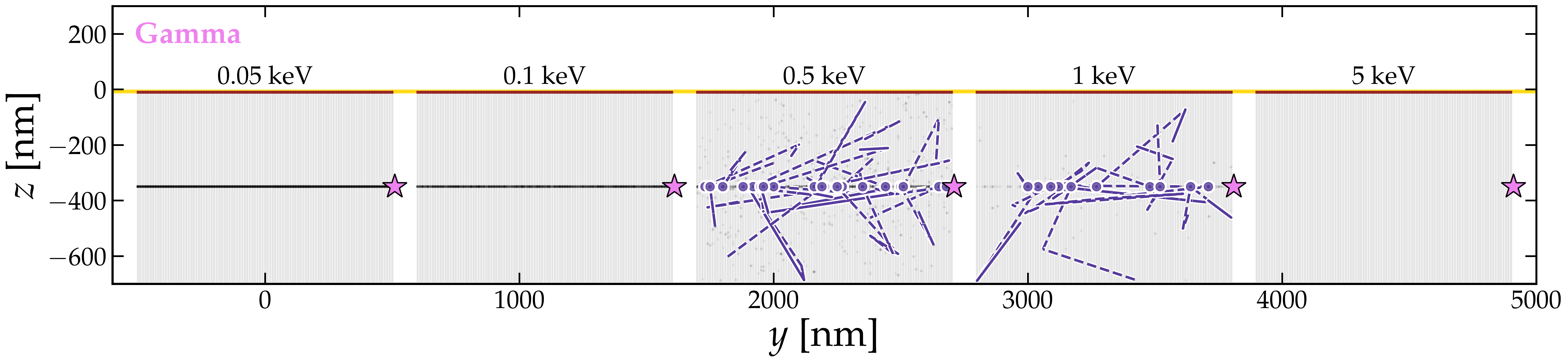}

    \caption{\label{fig:ExtraBreakage_sideon} Additional detailed images of strand breakage from side-on beams. Each row corresponds to a different beam particle, and has five panels for different beam energies. The thick lines connected by circles correspond to the primary particle track, whereas the thinner dashed lines correspond to secondary particles generated by a primary interaction.}
\end{figure}

\begin{figure}[t]
    \includegraphics[trim = 0mm 15mm 0 0mm,clip,width=0.95\textwidth]{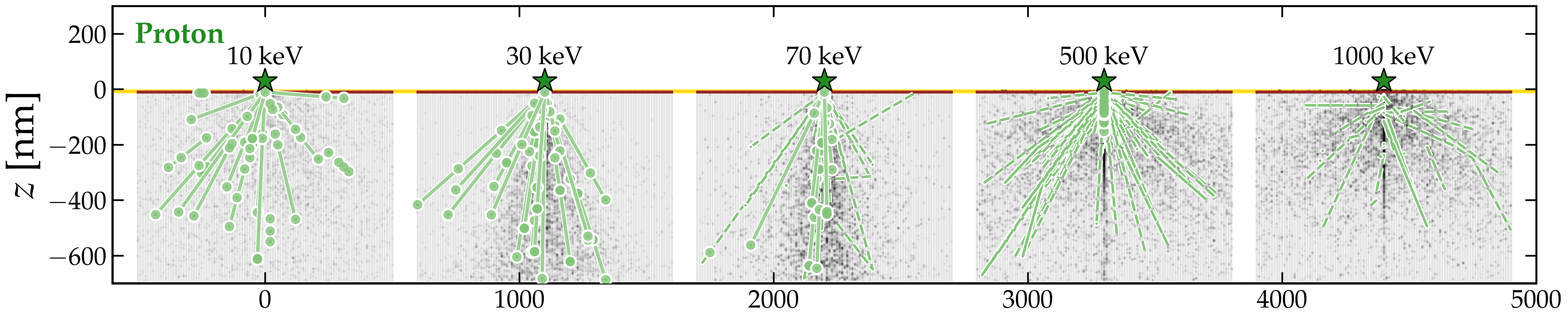}
    \includegraphics[trim = 0mm 15mm 0 0mm,clip,width=0.95\textwidth]{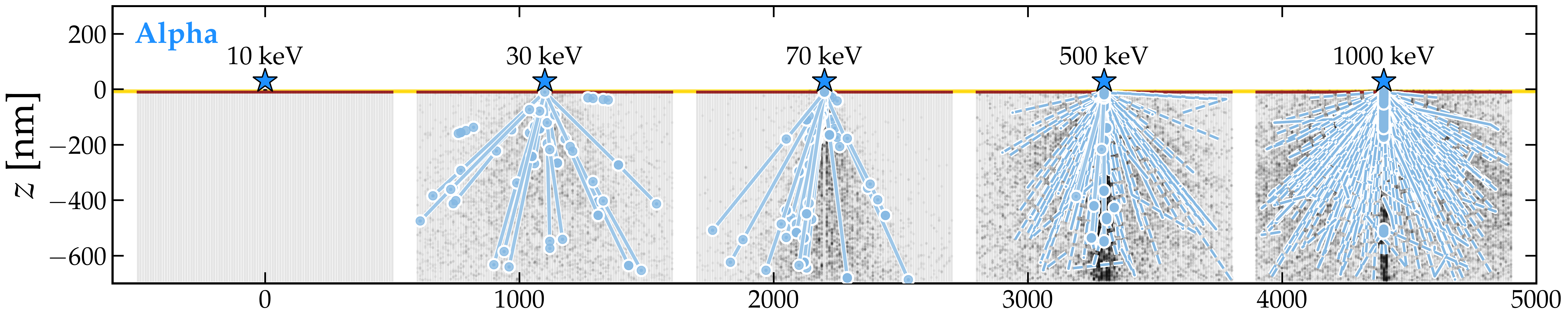}
    \includegraphics[trim = 0mm 15mm 0 0mm,clip,width=0.95\textwidth]{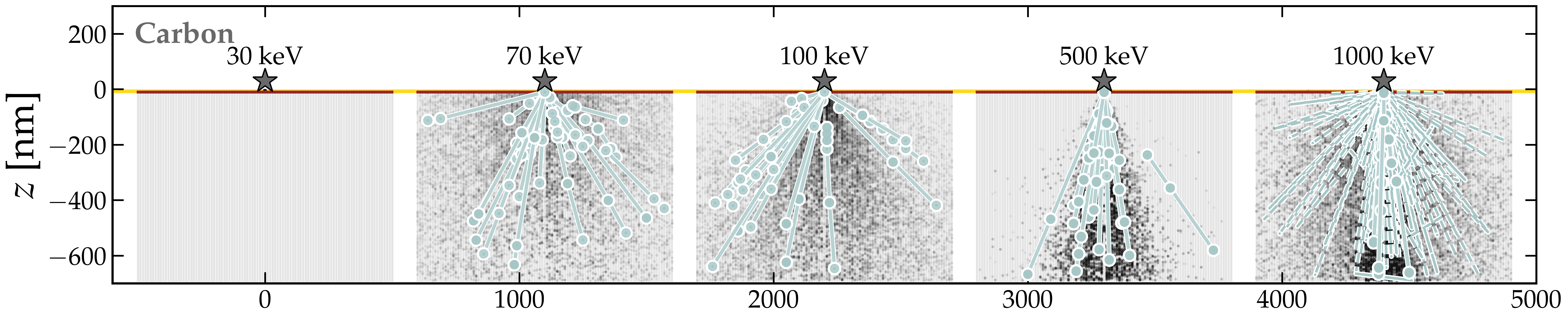}
    \includegraphics[trim = 0mm 15mm 0 0mm,clip,width=0.95\textwidth]{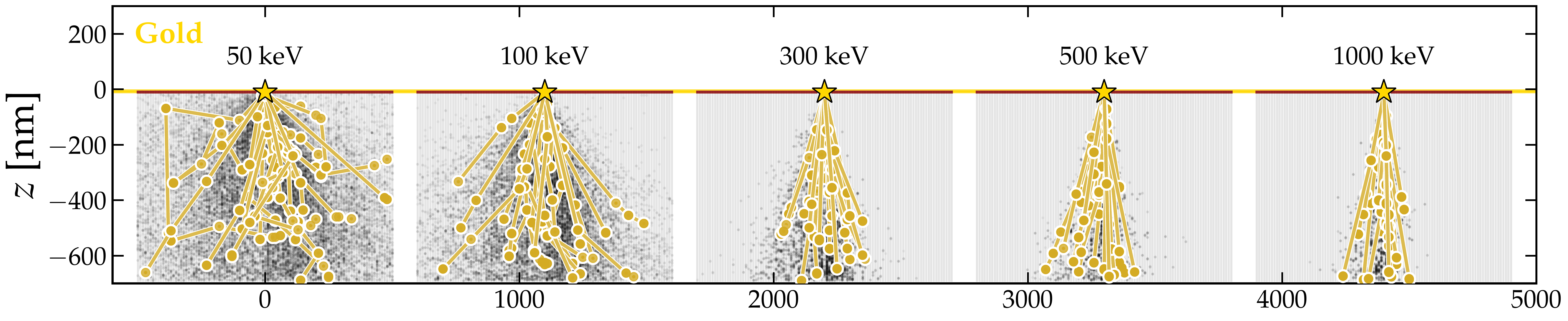}
        \includegraphics[trim = 0mm 15mm 0 0mm,clip,width=0.95\textwidth]{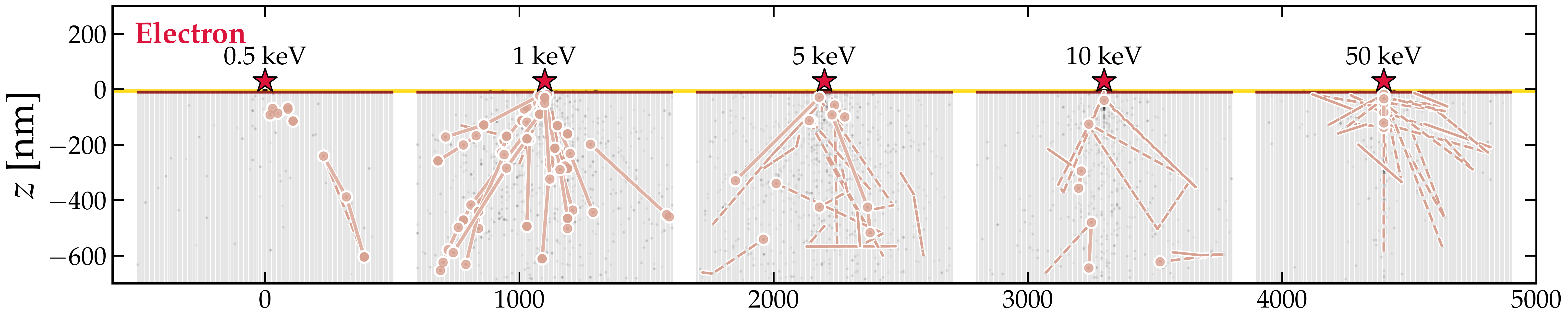}
    \includegraphics[width=0.95\textwidth]{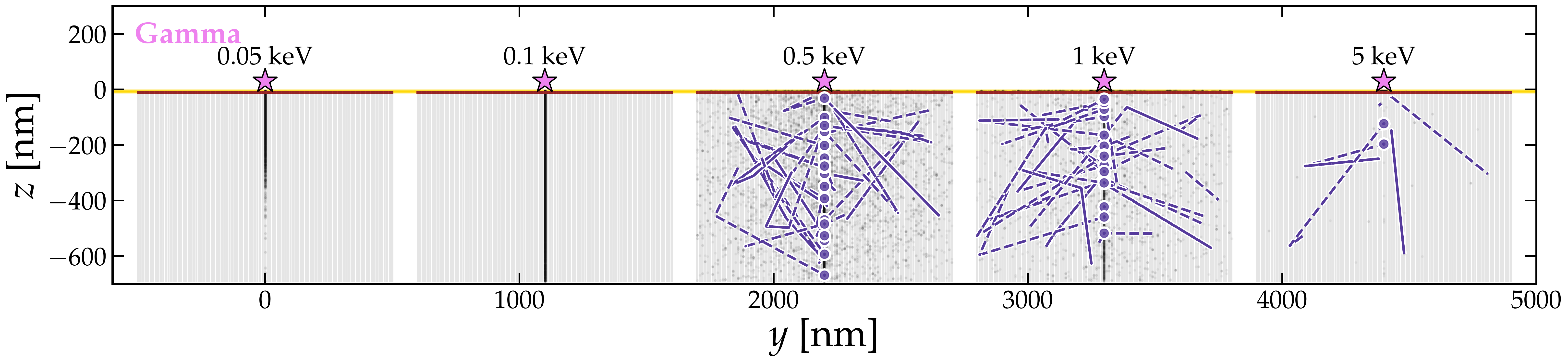}
    \caption{\label{fig:ExtraBreakage_topdown} Additional detailed images of strand breakage from top-down beams. Each row corresponds to a different beam particle, and has five panels for different beam energies. The thick lines connected by circles correspond to the primary particle track, whereas the thinner dashed lines correspond to secondary particles generated by a primary interaction. Those panels that show no interactions correspond to the cases where the incident particle is stopped inside the holder, or is not energetic enough to break a strand. The star denotes the entry position of the beam, note that we generate gold nuclei underneath the holder rather than above it.}
\end{figure}

\begin{figure}[t]
    \includegraphics[trim = 0mm 15mm 0 0mm,clip,width=0.95\textwidth]{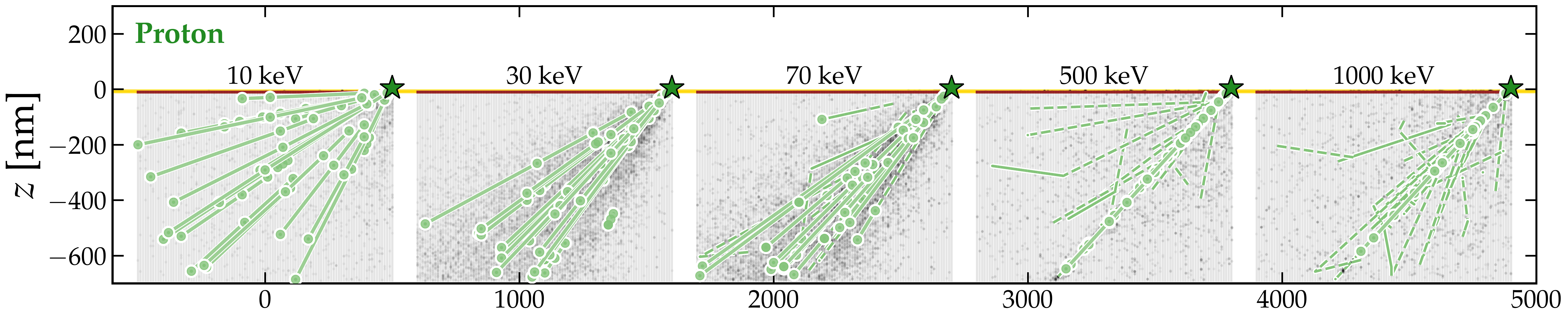}
    \includegraphics[trim = 0mm 15mm 0 0mm,clip,width=0.95\textwidth]{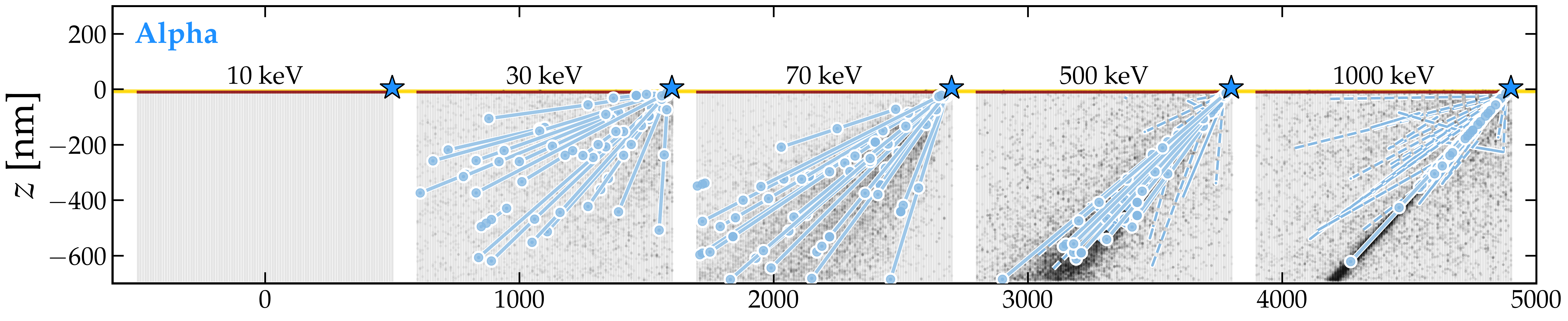}
    \includegraphics[trim = 0mm 15mm 0 0mm,clip,width=0.95\textwidth]{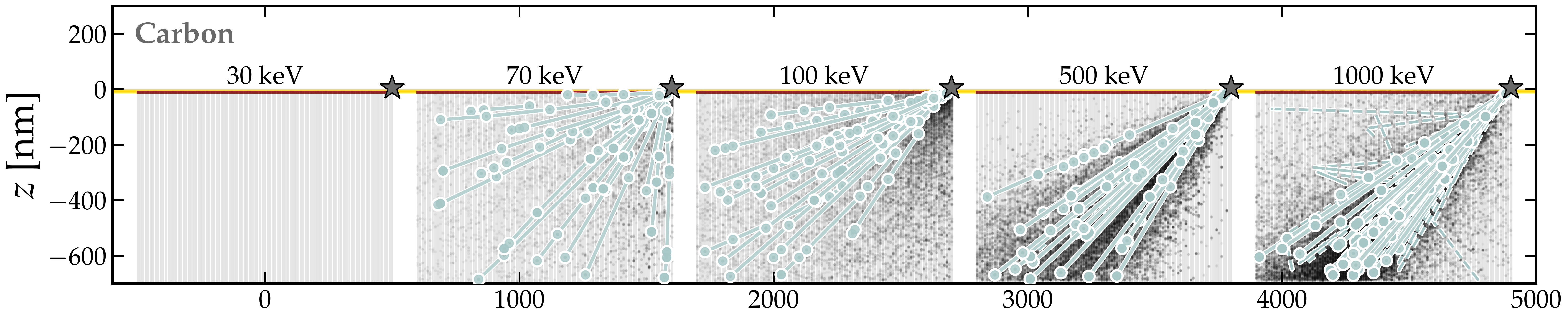}
    \includegraphics[trim = 0mm 15mm 0 0mm,clip,width=0.95\textwidth]{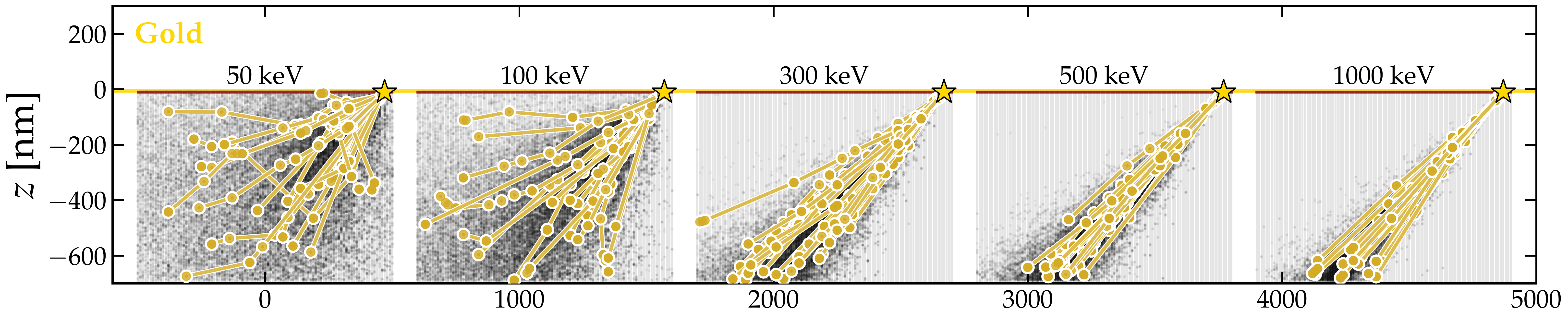}
    \includegraphics[trim = 0mm 15mm 0 0mm,clip,width=0.95\textwidth]{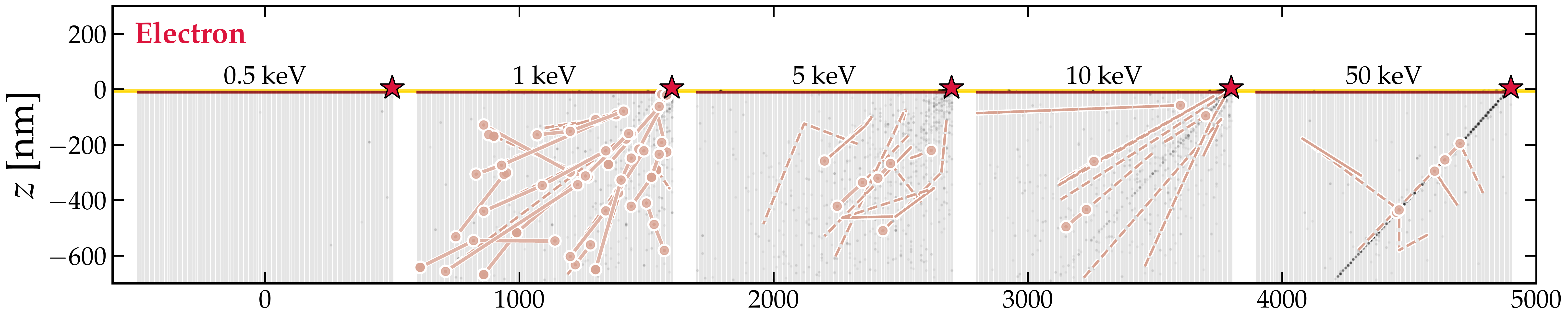}
        \includegraphics[width=0.95\textwidth]{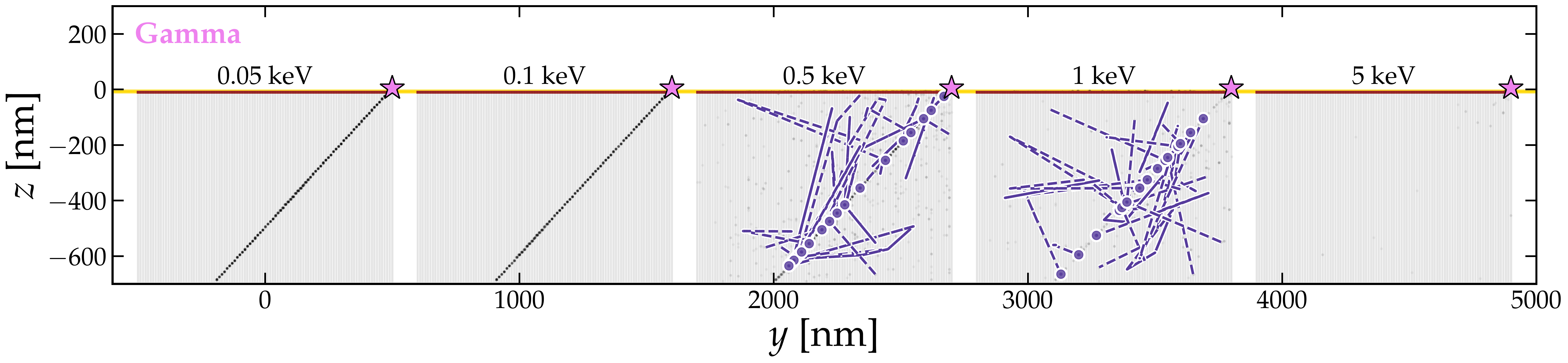}
    \caption{\label{fig:ExtraBreakage_angled} Additional detailed images of strand breakage from  beams angled at 45$^\circ$ to the holder. Each row corresponds to a different beam particle, and has five panels for different beam energies. The thick lines connected by circles correspond to the primary particle track, whereas the thinner dashed lines correspond to secondary particles generated by a primary interaction. The star denotes the entry position of the beam, note that we generate gold nuclei underneath the holder rather than above it.}
\end{figure}

In this appendix we display a larger set of a DNA breakage distributions for different particles, energies and orientations. We show only the $y-z$ projection of the distributions. These plots are here for completeness and to help further understand the behaviours described in the main text in Sec.~\ref{sec:results}. Figures~\ref{fig:ExtraBreakage_sideon},~\ref{fig:ExtraBreakage_topdown},~\ref{fig:ExtraBreakage_angled}, correspond to three beam angles, 0$^\circ$, 45$^\circ$ and 90$^\circ$, respectively.

\end{document}